\documentclass{aims} 
\usepackage{amsmath}
\usepackage{paralist}
\usepackage[misc]{ifsym}
\usepackage{epsfig} 
\usepackage{epstopdf} 
\usepackage[colorlinks=true]{hyperref}
\hypersetup{urlcolor=blue, citecolor=red}

\usepackage{amssymb} 
\usepackage{booktabs} 
\usepackage{bm} 
\usepackage{cleveref} 
\usepackage{float} 
\usepackage[acronym]{glossaries} 
\usepackage{makecell} 
\usepackage{mathtools} 
\usepackage{multirow} 
\usepackage{nicefrac} 
\usepackage{siunitx}
\usepackage{subcaption} 
\usepackage{todonotes} 
\usepackage{tikz} 
\usetikzlibrary{arrows,patterns,patterns.meta}   

\definecolor{custom-red}   {RGB}{215  48  39}
\definecolor{custom-orange}{RGB}{252 141  89}
\definecolor{custom-yellow}{RGB}{254 224 144}
\definecolor{custom-blue-1}{RGB}{224 243 248}
\definecolor{custom-blue-2}{RGB}{145 191 219}
\definecolor{custom-blue-3}{RGB}{ 69 117 180}

\allowdisplaybreaks

\def\currentvolume{X}
 \def\currentissue{X}
  \def\currentyear{200X}
   \def\currentmonth{XX}
    \def\ppages{X-XX}
     \def\DOI{10.3934/xx.xxxxxxx}

\theoremstyle{definition}

\newacronym{bc}{BC}{boundary condition}
\newacronym{cfd}{CFD}{Computational Fluid Dynamics}
\newacronym{dd}{DD}{domain decomposition}
\newacronym{dpinn}{DPINN}{Distributed Physics-Informed Neural Network}
\newacronym{fem}{FEM}{Finite Element Method}
\newacronym{ibc}{IBC}{initial and boundary condition}
\newacronym{ml}{ML}{Machine Learning}
\newacronym{nn}{NN}{Neural Network}
\newacronym{ode}{ODE}{ordinary differential equation}
\newacronym{pde}{PDE}{partial differential equation}
\newacronym{pinn}{PINN}{Physics-Informed Neural Network}
\newacronym{rans}{RANS}{Reynolds-Averaged Navier-Stokes}
\newacronym{rom}{ROM}{Reduced Order Model}
\newacronym{str}{STR}{stirred tank reactor}
\newacronym{xpinn}{XPINN}{Extended Physics-Informed Neural Network}

\title[A Model Hierarchy for Flows in Stirred Tanks with PINNs]
{A Model Hierarchy\\for Predicting the Flow in Stirred Tanks\\with Physics-Informed Neural Networks} 

\author[V. Trávníková, D. Wolff, N. Dirkes, S. Elgeti, E. von Lieres, and M. Behr]{}



\thanks{Corresponding author: Veronika Trávníková}
\thanks{$^*$Equal contribution}

\begin{document}
\maketitle

\centerline{\scshape
Veronika Trávníková$^{{\href{mailto:travnikova@cats.rwth-aachen.de}{\textrm{\Letter}}}*1}$, 
Daniel Wolff$^{{\href{mailto:d.wolff@unibw.de}{\textrm{\Letter}}}*1,2}$, 
Nico Dirkes$^{{\href{mailto:dirkes@aices.rwth-aachen.de}{\textrm{\Letter}}}1}$,}
\centerline{\scshape
Stefanie Elgeti$^{{\href{mailto:stefanie.elgeti@tuwien.ac.at}{\textrm{\Letter}}}3}$,
Eric von Lieres$^{{\href{mailto:e.von.lieres@fz-juelich.de}{\textrm{\Letter}}}4,5}$, 
and Marek Behr$^{{\href{mailto:behr@cats.rwth-aachen.de}{\textrm{\Letter}}}1}$}

\medskip

{\footnotesize
 \centerline{$^1$Chair for Computational Analysis of Technical Systems, RWTH Aachen University, Germany}
} 

\medskip

{\footnotesize
 \centerline{$^2$Institute for Mathematics and Computer-Based Simulation, University of the Bundeswehr Munich, Germany}
} 

\medskip

{\footnotesize
 \centerline{$^3$Institute for Lightweight Design and Structural Biomechanics, TU Wien, Austria}
}

\medskip

{\footnotesize
 \centerline{$^4$Institute of Bio- and Geosciences 1: Biotechnology, Forschungszentrum Jülich, Germany}
}

\medskip

{\footnotesize
 \centerline{$^5$Computational Systems Biotechnology, RWTH Aachen University, 52074 Aachen, Germany}
}

\bigskip

 \centerline{(Communicated by Handling Editor)}


\begin{abstract}
This paper explores the potential of \glspl{pinn} to serve as \glspl{rom} for simulating the flow field within \glspl{str}. We solve the two-dimensional stationary Navier-Stokes equations within a geometrically intricate
domain and explore methodologies that allow us to integrate additional physical insights into the model. These approaches include imposing the Dirichlet \glspl{bc} strongly and employing \gls{dd}, with both overlapping and non-overlapping subdomains. We adapt the \gls{xpinn} approach to solve different sets of equations in distinct subdomains based on the diverse flow characteristics present in each region. Our exploration results in a hierarchy of models spanning various levels of complexity, where the best models exhibit $\ell_1$ prediction errors of less than \qty{1}{\percent} for both pressure and velocity. To illustrate the reproducibility of our approach, we track the errors over repeated independent training runs of the best identified model and show its reliability. Subsequently, by incorporating the stirring rate as a parametric input, we develop a fast-to-evaluate model of the flow capable of interpolating across a wide range of Reynolds numbers. Although we exclusively restrict ourselves to \glspl{str} in this work, we conclude that the steps taken to obtain the presented model hierarchy can be transferred to other applications.

\end{abstract}


\section{Introduction}
\label{sec:introduction}
\Acrfullpl{str} play a central role in chemical process industry and biotechnology. A typical \gls{str} contains one or more impellers affixed to a shaft as well as baffles mounted to the reactor wall. A diverse set of parameters, including tank and impeller size, stirring rate, as well as the number of baffles, offer flexibility and control over the \gls{str}'s performance. However, these factors also pose significant challenges in designing and scale-up (or scale-down) of such reactors. Therefore, it is essential to understand the quantitative influence of these parameters on design objectives. As measurements are often prohibitively complicated, it is difficult to gain insight into the conditions inside these reactors. Models of \glspl{str} serve a dual purpose: They can help reduce the need for extensive experimental studies during process design and scale-up, while also enhancing our understanding of the conditions within the reactor. In industry, \gls{cfd} is already widely employed for this task. Nevertheless, high-fidelity simulations come at a significant computational cost, particularly in scenarios where the same model needs to be evaluated repeatedly with different parameters.

An example of such a scenario is the application of \glspl{str} as bioreactors in bioprocesses. A bioreactor typically contains a suspension of cells that are fed nutrients and, in turn, catalyze metabolic processes to produce desired products, such as antibiotics or antibodies. When modeling bioprocesses, two fundamental phenomena come into play: the hydrodynamics, which pertains to the flow field inside the reactor, and the kinetics of the biochemical reaction. These phenomena are closely interconnected, as heterogeneities in nutrient concentration lead to variations in intra-cellular composition among cells. Cells move within the tank during mixing, and even two cells in the same location at a given time may exist in different states. Consequently, a fast-to-evaluate flow model is essential for coupling with the mass transfer model \cite{Haringa2017}.

This challenge encourages the development of computationally efficient \acrfullpl{rom} that can provide fast approximate solutions for process design. Recent advancements in deep learning have sparked interest in employing data-driven approaches to address this need, particularly due to the advantage of offline pretraining and subsequent quick model evaluation offered by \gls{ml}. \gls{ml} has been employed to construct \glspl{rom}, for example, by utilizing autoencoders -- artificial neural networks that learn the mapping from high-dimensional input to a low-dimensional space \cite{YuYanGuo2019}. However, acquiring a sufficiently large training dataset for our specific application case proves prohibitively costly. Consequently, we aim to leverage our insights into the underlying physics of the problem to reduce or potentially eliminate the need for training data. 

\acrfullpl{pinn} integrate the governing equations of the problem into the loss function of the \gls{nn}, making them a natural fit for our objectives. Since the seminal paper of Raissi et al. (2019) \cite{Raissi2019}, \glspl{pinn} have been successfully applied to problems across diverse domains, including solid mechanics \cite{laiStructuralIdentificationPhysicsinformed2021, Haghighat2021, sahinSolvingForwardInverse2023}, molecular dynamics \cite{IslamMD2021}, chemical reaction kinetics \cite{adebarScientificDeepMachine2024}, the wave equation \cite{moseley2020wave}, hemodynamics \cite{Kissas2020, Sun2020, RaissiHFM2020}, cardiac activation mapping \cite{sahlicostabalPhysicsInformedNeuralNetworks2020}, and various other fields \cite{Cuomo2022}. Furthermore, \glspl{pinn} have been applied to address numerous problems in fluid mechanics \cite{JinNSFnets2021, ChengZhang2021, Jagtap2022, tillmannShapeoptimizationExtrusiondiesParameterized2023,FaroughiReview2024, zhouAdvancingFluidDynamics2024}. Notably, they have been employed to reconstruct the 3D wake flow past a cylinder using data from a few cross-planes \cite{Cai2021} or as closure for \gls{rans} equations \cite{Eivazi2022}. The approach presented in this paper extends beyond embedding the governing equations into the model training. We incorporate additional insights by implementing strong \acrfullpl{bc} \cite{Sheng2021LFF} with a lifting function, aligned with an estimated velocity profile. Additionally, we perform \acrfull{dd}, solving different sets of equations in different subdomains according to the distinct flow characteristics present in each respective region.

The rest of the paper is structured as follows: \Cref{sec:methods} briefly introduces the reader to the mathematical basics of \glspl{nn} and \glspl{pinn}. Since the field of \glspl{pinn} is rapidly evolving, we will only focus on presenting the developments that are of immediate relevance to our work such as \acrfullpl{xpinn} and strong \gls{bc} imposition. After that, we present the application to which we are employing our method in \Cref{sec:model}. We explain the model problem that we will consider in this paper in detail, which includes stating our assumptions, the fundamental physical conservation laws that govern the flow in the \gls{str} and the employed \glspl{bc}. We conclude the theory part of the paper by elaborating on the validation strategy for the \gls{pinn} models. The following \Cref{sec:nonparameterized,sec:parameterized} then contain the main findings of this work. We start with presenting a model hierarchy for a simpler case, where we only train \glspl{pinn} to predict the flow field for one fixed Reynolds number in \Cref{sec:nonparameterized}. Here, we detail how we successively tested different approaches to increase the predictive accuracy of the reduced simulation models. Since we are interested in reliable \glspl{rom} for multi-query scenarios, we then transfer these findings in \Cref{sec:parameterized} to predict the flow field inside our \gls{str} geometry for different impeller velocities. Finally, \Cref{sec:conclusion} concludes this work by summarizing the main results and providing an outlook on future work.


\section{Methods}
\label{sec:methods}
In this section, we briefly outline the basics of the methods applied in this paper. Consequently, we first recapitulate the mathematical definition of a fully-connected feed-forward \gls{nn} --- the \gls{nn} architecture to which we restrict ourselves within this paper --- before we introduce our \gls{pinn}-related notation. To keep both sections concise, we will only mention the aspects that are most important for this paper and refer to the literature for a more in-depth discussion of the respective methods.


\subsection{Fully-connected feed-forward \glspl{nn}}
\label{subsec:methods:fcffnn}
In this work, we only consider fully-connected feed-forward \glspl{nn}, that is, every neuron of the preceding layer is connected to every neuron of the following layer and all connections are directed forward, i.e., information is only passed from one layer onward to the next. Mathematically, a \gls{nn} mapping from an input space $\mathcal{X}$ to an output space $\mathcal{U}$ and consisting of $N_{L}$ layers can be formalized as 
\begin{align}
    \label{eq:methods:fcffnn:network}
    \bm{\mathcal{NN}} : \mathcal{X} \to \mathcal{U} \;,\; \bm{x}\mapsto\bm{\mathcal{NN}}(\bm{x};\bm{\theta})=\left(\bm{f}^{(N_{L})} \circ \hdots \circ \bm{f}^{(1)}\right)(\bm{x};\bm{\theta}) \; ,
\end{align}
where each layer is described by a function
\begin{align}
    \label{eq:methods:fcffnn:layer}
    \bm{f}^{(l)}\left(\bm{x};\bm{\theta}^{(l)}\right) &= \sigma^{(l)}\left(\bm{W}^{(l)}\bm{x}+\bm{b}^{(l)}\right) \quad \forall \; l\in\{1,\hdots,N_{L}\} \; .
\end{align}
Here, $\bm{\theta}^{(l)}=\{\bm{W}^{(l)},\bm{b}^{(l)}\}$ denotes the trainable parameters, namely the layer's weights and biases, while $\sigma$ denotes the activation function acting element-wise on its input. In principle, one can employ different activation functions after every layer. For the rest of this paper, however, we restrict ourselves to the same activation after all hidden layers and a linear activation function in the last layer, that is $\sigma^{(l)}(\bm{z})=\sigma(\bm{z}) \;\forall\;l\in\{1,\hdots,N_{L-1}\}$ and $\sigma^{(N_L)}(\bm{z})=\bm{z}$. 
For the complete network as defined in \Cref{eq:methods:fcffnn:network}, $\bm{\theta}=\{\bm{\theta}^{(1)},\hdots,\bm{\theta}^{(N_L)}\}$ denotes the collection of layer-wise parameters and $\bm{x}\in\mathcal{X}$ the input vector. 

In the following, we will employ fully-connected feed-forward \glspl{nn} only for regression tasks, i.e., $\mathcal{X}\subseteq\mathbb{R}^{N_{\text{in}}}$ and $\mathcal{U}\subseteq\mathbb{R}^{N_{\text{out}}}$.


\subsection{\acrfullpl{pinn}}
\label{subsec:methods:pinn}
The concept of \glspl{pinn} was already proposed in the last century \cite{Lagaris1998,Psichogios1992} but received again attention in recent years \cite{Raissi2019} after the advances in computational resources and broad availability of efficient algorithmic differentiation implementations made the approach more viable.
The basic idea of \glspl{pinn} is to let a \gls{nn} (e.g., as introduced in \Cref{eq:methods:fcffnn:network}) approximate the unknown solution field of a (partial) differential equation. Let us consider the following generic formulation of a \gls{pde} (system)
\begin{subequations}
\label{eq:methods:pinn:pdeandbc}
\begin{align}
    \bm{\mathcal{R}}[\bm{u}] &= \bm{0} \quad \text{in} \; \Omega \; ,\\
    \bm{\mathcal{B}}[\bm{u}] &= \bm{0} \quad \text{on} \; \Gamma \; ,
\end{align}
\end{subequations}
where $\bm{\mathcal{R}}[\,\cdot\,]$ denotes the \gls{pde} and $\bm{\mathcal{B}}[\,\cdot\,]$ the \gls{ibc} operator, both formulated as residuals. In the general case of \gls{pde} systems, $\bm{\mathcal{R}}[\,\cdot\,]$ and $\bm{\mathcal{B}}[\,\cdot\,]$ are usually vectors. $\bm{u}$ corresponds to the (unknown) \gls{pde} solution which will now be approximated by a \gls{nn}
\begin{align}
    \label{eq:methods:pinn:approximation}
    \bm{u}(\bm{x}) \approx \bm{u}_{\bm{\theta}}(\bm{x}) \coloneqq \bm{\mathcal{NN}}(\bm{x};\bm{\theta}) \; .
\end{align}
During training, this \gls{nn} first predicts the approximated \gls{pde} solution at (randomly) selected points of the domain $\Omega$ and the boundary $\Gamma=\partial\Omega$. The deviation from the true solution is then subsequently quantified by evaluating the residuals of \Cref{eq:methods:pinn:pdeandbc} on these predictions. This yields the following \emph{physics-informed} loss function for the training:
\begin{subequations}
\label{eq:methods:pinn:physicslosses}
\begin{align}
    \label{eq:methods:pinn:physicsloss}
    \mathcal{L}^{\text{PINN}}_{\text{phys}}(\bm{\theta};\bm{X}) &= \bm{\alpha}_{\text{PDE}} \cdot \bm{\mathcal{L}}_{\text{PDE}}(\bm{\theta}; \bm{X}_{\text{PDE}}) + \bm{\alpha}_{\text{IBC}} \cdot \bm{\mathcal{L}}_{\text{IBC}}(\bm{\theta}; \bm{X}_{\text{IBC}}) \; ,
\end{align}
with
\begin{align}
    \label{eq:methods:pinn:physicslosses:pdes}
    \bm{\mathcal{L}}_{\text{PDE}}(\bm{\theta}; \bm{X}_{\text{PDE}}) &= \frac{1}{N_{\text{PDE}}} \sum_{i=1}^{N_{\text{PDE}}} \left(\bm{\mathcal{R}}\left[\bm{u}_{\bm{\theta}}\left(\bm{x}^{(i)}_{\text{PDE}}\right)\right]\right)^2 \; , \\
    \label{eq:methods:pinn:physicslosses:bcs}
    \bm{\mathcal{L}}_{\text{IBC}}(\bm{\theta}; \bm{X}_{\text{IBC}}) &= \frac{1}{N_{\text{IBC}}} \sum_{i=1}^{N_{\text{IBC}}} \left(\bm{\mathcal{B}}\left[\bm{u}_{\bm{\theta}}\left(\bm{x}^{(i)}_{\text{IBC}}\right)\right]\right)^2 \; .
\end{align}
\end{subequations}
In the case of vector-valued operators $\bm{\mathcal{R}}[\,\cdot\,]$ and $\bm{\mathcal{B}}[\,\cdot\,]$, the square operation in our notation is to be understood component-wise so that $\bm{\mathcal{L}}_{\text{PDE}}$ and $\bm{\mathcal{L}}_{\text{IBC}}$ are again vectors. Moreover, $\bm{X}=\{\bm{X}_{\text{PDE}},\bm{X}_{\text{IBC}}\}$ denotes the set of evaluation points --- also referred to as \emph{collocation points} --- on which the \gls{pde} as well as the (initial and) \glspl{bc} are evaluated. $\bm{X}_{\text{PDE}}=\left\{\bm{x}^{(1)}_{\text{PDE}},\hdots,\bm{x}^{(N_{\text{PDE}})}_{\text{PDE}}\right\}$ and $\bm{X}_{\text{IBC}}=\left\{\bm{x}^{(1)}_{\text{IBC}},\hdots,\bm{x}^{(N_{\text{IBC}})}_{\text{IBC}}\right\}$ represent the collocation points in the domain and on the boundary respectively. $\bm{\alpha}_{\text{PDE}}$ and $\bm{\alpha}_{\text{IBC}}$ are scaling factors that allow to weigh the individual contributions of the physical constraints differently.

\subsubsection*{Selecting appropriate loss scaling factors.} A great deal of research has been dedicated to the choice of these scaling factors, as they have been found to severely impact the convergence of the training as well as the accuracy of the prediction of \glspl{pinn} \cite{Wang2021LossScaleAnnealing}. Several methods have been proposed to choose these scaling factors appropriately. Most of these methods involve dynamic updates during training. For linear and well-posed \gls{pde} problems, van der Meer et al. (2022) \cite{VanDerMeer2022OptimalLossWeighting} derived optimal loss scaling factors. However, that requires an analytical solution which is usually not available in engineering applications. In addition, they proposed a heuristic that alleviates this requirement. In \cite{Li2022BiLevelOptiWeights}, Li et al. (2022) proposed a bi-level optimization algorithm where the loss scales are adapted by gradient descent steps. Another method to adapt the loss scales was proposed by Wang et al. (2021) \cite{Wang2021LossScaleAnnealing}, which is based on the magnitude of the gradients of the loss components. In their follow-up work \cite{Wang2022NTK}, they proposed another update mechanism, leveraging insights from neural tangent kernel theory.

\subsubsection*{A remark about data}
Although the general idea of \glspl{pinn} is to overcome the need for training data --- which is often scarce in engineering applications --- available data can be integrated into the framework presented above. To that end, \Cref{eq:methods:pinn:physicsloss} can be modified as follows:
\begin{subequations}
\label{eq:methods:pinn:physicsdataloss}
\begin{align}
     \mathcal{L}^{\text{PINN}}_{\text{phys+data}}(\bm{\theta};\bm{X}) &= \mathcal{L}^{\text{PINN}}_{\text{phys}}(\bm{\theta};\{\bm{X}_{\text{PDE}},\bm{X}_{\text{IBC}}\}) + 
     \bm{\alpha}_{\text{data}} \cdot \bm{\mathcal{L}}_{\text{data}}(\bm{\theta}; \bm{X}_{\text{data}}) \; ,
\end{align}
where
\begin{align}
    \bm{\mathcal{L}}_{\text{data}}(\bm{\theta}; \bm{X}_{\text{data}}) &= \frac{1}{N_{\text{data}}} \sum_{i=1}^{N_{\text{data}}} \left( \bm{u}_{\bm{\theta}}\left(\bm{x}^{(i)}_{\text{data}}\right)-\bm{u}^{(i)}_{\text{data}} \right)^2 \; .
\end{align}
\end{subequations}
Here, $\bm{X}_{\text{data}}=\left\{\left(\bm{x}^{(1)}_{\text{data}},\bm{u}^{(1)}_{\text{data}}\right),\hdots,\left(\bm{x}^{(N_{\text{data}})}_{\text{data}},\bm{u}^{(N_{\text{data}})}_{\text{data}}\right)\right\}$ represents a training set with high-fidelity information on the solution $\bm{u}^{(i)}_{\text{data}}$ at input locations $\bm{x}^{(i)}_{\text{data}}$.


\subsection{Advanced \gls{pinn} architectures}
\label{subsec:methods:advanced}

The preceding section covered the fundamental \gls{pinn} architecture. Nevertheless, at the time of writing, a variety of more complex architectures have been proposed \cite{Jagtap2020CPINNs,Kharazmi2019vPINNs,Mojgani2023LPINNs,Moseley2023FBPINNs,Wang2021LossScaleAnnealing}, which have been tailored to overcome known drawbacks of \glspl{pinn} or to improve the predictive capabilities beyond those of plain baseline \glspl{pinn}. Due to their relevance for the work of this paper, we will briefly outline the idea of \glspl{xpinn} \cite{Jagtap2020XPINNs} in the following. In the literature, a similar approach has been reported under the name \glspl{dpinn} \cite{Dwivedi2021}.

The general idea is motivated by \gls{dd} approaches: The computational domain $\Omega$ is decomposed into $N_d$ non-overlapping subdomains $\Omega_k$ such that $\bigcup_{k=1}^{N_d} \Omega_k=\Omega$, with $\Omega_i\cap\Omega_j=\Gamma_{ij} \; \forall \; i,j\leq N_d \wedge i \neq j$. Within each of these subdomains $\Omega_k$, a different \gls{pinn} $\bm{u}_{\bm{\theta}^{(k)}}$ is trained. This approach allows to adapt the architecture of the network to the structure of the solution: In regions of the domain where smooth solutions are expected, the networks can be narrower and shallower, while deeper and wider network architectures might be required in regions where the solution is more complex. The complete \gls{pde} solution is then assembled from the individual sub-networks' predictions in their specific regions, that is,
\begin{align}
    \bm{u}_{\bm{\theta}}(\bm{x}) &= \sum_{k=1}^{N_d} \bm{u}_{\bm{\theta}^{(k)}}(\bm{x}) \mathbb{I}_{\Omega_k}(\bm{x}) \; ,
\end{align}
with the indicator function
\begin{align}
    \mathbb{I}_{\Omega_k}(\bm{x}) &= 
    \begin{cases}
        1 & \bm{x} \in \Omega_k \setminus \Gamma_k \\
        \frac{1}{2} & \bm{x} \in \Gamma_k \\
        0 & \bm{x} \notin \Omega_k
    \end{cases}
    \; .
\end{align}
All networks are trained simultaneously. Efficient methods for parallelizing the training of the $N_d$ networks have been proposed, e.g., in \cite{Shukla2021PPINNs}. The predictions of the different sub-networks are coupled by augmenting \Cref{eq:methods:pinn:physicsloss} with additional terms representing the equality of the residuals and average values across sub-domain boundaries as well as continuity constraints on the solution values and their derivatives (if needed).

\subsection{Input and output scaling}
\label{subsec:methods:inputoutputscaling}

For all models presented in \Cref{sec:nonparameterized,sec:parameterized}, we normalize inputs and outputs to improve the learning process. Essentially, we non-dimensionalize inputs and outputs as suggested in literature \cite{Wang2023ExpertGuide} by adding fixed (i.e., not learned) feature transformation layers to the network that transform the inputs to be roughly contained in the interval $[-1,1]^{N_\text{in}}$ and rescale the outputs from $[-1,1]^{N_\text{out}}$ to the actual dimensional quantities. Mathematically, we augment \Cref{eq:methods:pinn:approximation} by additionally composing it with a pre-processing function $\bm{f}_\text{pre}$ and a post-processing function $\bm{f}_\text{post}$, so that
\begin{align}
    \label{eq:methods:inputoutputscaling:ansatz}
    \bm{u}_{\bm{\theta}}(\bm{x}) &= \bm{f}_\text{post}(\bm{x}) \circ \bm{\mathcal{NN}}(\bm{x};\bm{\theta}) \circ \bm{f}_\text{pre}(\bm{x}) \; .
\end{align}
The pre- and post-processing functions employed in our work differ among the different models and will be reported in the respective sections.


\subsection{Strong imposition of Dirichlet \glspl{bc}}
\label{subsec:methods:strongDirichlet}

So far, in the previous sections we have only discussed the case where the \glspl{bc} of the \gls{pde} problem \Cref{eq:methods:pinn:pdeandbc} are enforced weakly, that is, via a penalty term in the loss function (c.f. \Cref{eq:methods:pinn:physicslosses:bcs}). However, research has been dedicated to imposing Dirichlet \glspl{bc} of the form $\bm{u}(\bm{x})=\bm{h}(\bm{x}) \; \text{on} \; \Gamma_D$ strongly, similar to what is often done in the \gls{fem}. Mathematically, the strong imposition of Dirichlet boundaries takes the form of a post-processing operation, as discussed in the previous section, i.e., the ansatz \Cref{eq:methods:pinn:approximation} is modified as follows:
\begin{align}
    \label{eq:methods:pinn:approximation:strong}
    \bm{u}_{\bm{\theta}}(\bm{x}) &= \bm{h}(\bm{x}) + \bm{g}(\bm{x}) \odot \bm{\mathcal{NN}}(\bm{x};\bm{\theta}) \; ,
\end{align}
where $\bm{h}(\bm{x})$ is the (known) Dirichlet \gls{bc} on the Dirichlet part of the boundary $\Gamma_D\subseteq\Gamma$ and $\bm{g}(\bm{x})$ a function that depends on the geometry of the computational domain $\Omega$ with the properties
\begin{align}
    \begin{cases}
        \bm{g}(\bm{x}) = \bm{0} & \text{on} \; \Gamma_D \\
        \bm{g}(\bm{x}) > \bm{0} & \text{in} \; \Omega \setminus \Gamma_D
    \end{cases}
    \; .
\end{align}
The symbol $\odot$ in \Cref{eq:methods:pinn:approximation:strong} symbolizes the component-wise multiplication of two vectors.

In the literature, different approaches have been proposed of how to construct the positive semi-definite function $\bm{g}$ which vanishes on the Dirichlet boundary. For simple rectangular geometries, Lagari et al. (2020) \cite{Lagari2020} describe a construction using polynomials and exponential functions that generalizes to arbitrary dimensions and can also be applied to Neumann and Robin conditions. For more complex geometries, an algorithm has been proposed in \cite{Sheng2021LFF} that constructs $\bm{g}$ through a radial-basis function interpolation of point samples generated on the domain boundary. 


\section{Model}
\label{sec:model}

In this section, we introduce the simulation model that serves as the basis for the construction of the corresponding \gls{pinn}-based \glspl{rom} proposed in this paper. We discuss the governing equations and explain the test case that will be investigated in the following \Cref{sec:nonparameterized,sec:parameterized} before briefly commenting on the classical numerical solution approach via the \gls{fem} which we used to generate the validation results for our models.


\subsection{Assumptions}
\label{subsec:model:assumptions}

As mentioned in the introduction, modeling the flow field inside \glspl{str} poses a number of challenges. For developing \gls{pinn}-based \glspl{rom}, we consequently make some simplifying assumptions to reduce the complexity of the learning task, especially since we are interested in approaches that learn solutions of reasonable accuracy without the help of high-fidelity data. Therefore, we assume
\begin{itemize}
    \item an incompressible, Newtonian fluid, i.e., the density $\rho$ and viscosity $\mu$ are constant,
    \item a single phase flow, i.e., we neglect inhomogeneities inside the vessel,
    \item and a steady flow, i.e., we are only interested in finding the solution for the flow field at a specific instance in time.
\end{itemize}
The last simplification enables us to circumvent the complexities associated with the moving domain that would arise from a rotating stirrer. 
While these simplifications shift the model away from realistic applications such as bioreactors, they allow us to conduct a feasibility study and identify the most promising methodologies. Moving forward, our aim is to apply these methodologies to models with increased complexity and explore more realistic scenarios.

Before we introduce the problem domain $\Omega$ that will be considered throughout \Cref{sec:nonparameterized,sec:parameterized}, we first briefly state the governing equations resulting from the assumptions above.


\subsection{Governing equations}
\label{subsec:model:goveq}

The mass and momentum balance inside the problem domain are given by the stationary, incompressible Navier-Stokes equations. The total \gls{pde} residual in \Cref{eq:methods:pinn:physicslosses:pdes} consists of the following components $\bm{\mathcal{R}}[\bm{u}]=(\bm{\mathcal{R}}_{\text{momentum}}[\bm{u}]^T,\mathcal{R}_{\text{mass}}[\bm{u}])^T$. We solve these for the unknown velocity vector $\bm{v}=(v_x,v_y)^T$ and the unknown pressure $p$, i.e., $\bm{u}=\left(\bm{v}^T,p\right)^T$. The residuals of the mass and momentum balance are given by
\begin{subequations}
\label{eq:model:governingequations}
\begin{align}
    \bm{\mathcal{R}}_{\text{momentum}}[\bm{u}] &= \rho (\bm{v}\cdot\bm{\nabla}) \bm{v} + \bm{\nabla} p - \mu \Delta \bm{v} \; , \\
    \mathcal{R}_{\text{mass}}[\bm{u}] &= \bm{\nabla}\cdot\bm{v} \; .
\end{align}
\end{subequations}
These conservation laws need to be equipped with suitable \glspl{bc} that depend on the problem domain, which we will introduce next.


\subsection{Problem domain and \acrlongpl{bc}}
\label{subsec:model:problemdomainbcs}

As modeling the flow inside a realistic three-dimensional geometry is a challenging task and the focus of this paper is on laying the ground-work by identifying promising methodologies to be translated into the development of a \gls{rom} for more realistic scenarios, we first consider the planar flow in a two-dimensional cross section through the stirrer plane (see \Cref{fig:model:domain:overview:3d}). The resulting computational domain is depicted in \Cref{fig:model:domain:overview:2d}.
\begin{figure}[ht]
    \centering
    \begin{subfigure}[b]{0.47\textwidth}
        \centering
\newlength{\gridlength}
\setlength{\gridlength}{15cm}

\begin{tikzpicture}[font=\Large, x=\gridlength, y=\gridlength]
    
    \draw[thick, gray] (-0.1,0) -- (-0.1,0.3);
    \draw[thick, gray] (0,0.3) ellipse (0.1 and 0.02);
    \draw[thick, gray] (0.1,0.3) -- (0.1,0);
    \draw[thick, gray] (0.1,0) arc[start angle=0, end angle=-180, x radius=0.1, y radius=0.02];
    
    \fill[thick, custom-red!10, draw=custom-red] (-0.1,0) -- (-0.1,0.225) -- (-0.085,0.225) -- (-0.085,0) -- cycle;
    \fill[thick, custom-red!10, draw=custom-red] (0.1,0) -- (0.1,0.225) -- (0.085,0.225) -- (0.085,0) -- cycle;
    
    \fill[thick, custom-orange!10, draw=custom-orange] (-0.03,0.07) -- (0,0.06) -- (0,0.05) -- (-0.03,0.06) -- cycle;
    \fill[thick, custom-orange!10, draw=custom-orange] (0,0.06) -- (0.03,0.07) -- (0.03,0.06) -- (0,0.05) -- cycle;
    \fill[thick, custom-orange!10, draw=custom-orange] (-0.03,0.05) -- (0,0.06) -- (0,0.05) -- (-0.03,0.04) -- cycle;
    \fill[thick, custom-orange!10, draw=custom-orange] (0,0.06) -- (0.03,0.05) -- (0.03,0.04) -- (0,0.05) -- cycle;
    \draw[thick, custom-orange] (0,0.3) -- (0,0.05);
    
    \fill[dashed, very thick,custom-blue-1, draw=custom-blue-3, opacity=0.5] (0,0.055) ellipse (0.1 and 0.02);
    
    \draw[thick, black, -{latex}] (-0.15,-0.05) -- (-0.15,0.025) node[above] {$z$};
    \draw[thick, black, -{latex}] (-0.15,-0.05) -- ++(0.05032,0.01677) node[right] {$y$};
    \draw[thick, black, -{latex}] (-0.15,-0.05) -- ++(0.05032,-0.01677) node[right] {$x$};

\end{tikzpicture}
        \caption{Simplified schematic of a three-dimensional \gls{str}. The considered two-dimensional cross section through the stirrer plane is indicated by the blue ellipse and depicted in more detail in \Cref{fig:model:domain:overview:2d}.}
        \label{fig:model:domain:overview:3d}
    \end{subfigure}
    \hfill
    \begin{subfigure}[b]{0.47\textwidth}
        \centering
\setlength{\gridlength}{20cm}

\begin{tikzpicture}[font=\Large, x=\gridlength, y=\gridlength]
    
    \fill[very thick, draw=custom-red, fill=custom-blue-1] (0,0) circle (0.1);
    \node[custom-blue-3] at (0,0.075) {$\Omega$};
    \node[custom-red] at (-0.08,-0.085) {$\Gamma_\text{wall}$};
    
    \begin{scope}
        \clip (0,0) circle (0.1);
        \def \sqrttwo {1.41421356237}
        \def \position {0.1/\sqrttwo}
        \foreach \x/\y/\d in {\position/\position/-45, \position/-\position/45, -\position/-\position/-45, -\position/\position/45} {
            \node[very thick, draw=custom-red, fill=custom-red!10, shape=rectangle, minimum width=0.005\gridlength, minimum height=0.03\gridlength, inner sep=0pt, anchor=center, rotate=\d] at (\x,\y) {};
        }
    \end{scope}

    \draw[thick, gray] (-0.04,0.02) -- (0.04,-0.02);
    \draw[thick, gray] (-0.02,-0.04) -- (0.02,0.04);
    \foreach \r in {0.02, 0.03, 0.04} {
        \draw[thick, gray, -{latex}] (\r,0) -- (\r,-0.5*\r);
        \draw[thick, gray, -{latex}] (0,\r) -- (0.5*\r,\r);
        \draw[thick, gray, -{latex}] (-\r,0) -- (-\r,0.5*\r);
        \draw[thick, gray, -{latex}] (0,-\r) -- (-0.5*\r,-\r);
    }
    
    \foreach \r in {-0.04, 0.04} {
        \draw[very thick, custom-orange] (0,0) -- (0,\r);
        \draw[very thick, custom-orange] (0,0) -- (\r,0);
    }
    \node[custom-orange] at (0.04,-0.04) {$\Gamma_\text{stirrer}$};
    
    \draw[thick, -{latex}] (0,-0.07) to [out=180,in=270, looseness=1] (-0.07,0);
    \node at (-0.07,0.015) {$\omega$};
    
    \draw[thick, black, -{latex}] (-0.125,-0.125) -- (-0.125,-0.075) node[above] {$y$};
    \draw[thick, black, -{latex}] (-0.125,-0.125) -- (-0.075,-0.125) node[right] {$x$};
    \draw[thick, black, -{latex}] (-0.125,-0.125) circle (1pt);
    \draw[thick, black, -{latex}] (-0.125,-0.125) circle (4pt) node[below left=0.003] {$z$};
    
\end{tikzpicture}
        \caption{Schematic of the bioreactor-inspired problem domain $\Omega$. The stirrer is assumed to be rotating with a constant angular velocity $\omega$, resulting in a linear velocity profile on the stirrer blades $\Gamma_\text{stirrer}$.}
        \label{fig:model:domain:overview:2d}
    \end{subfigure}
    \caption{Schematic depiction of a three-dimensional \gls{str} geometry (A) as well as the simplified two-dimensional problem domain considered in the rest of this paper (B).}
    \label{fig:model:domain:overview:overview}
\end{figure}

This choice of the computational domain can be understood as follows: Our frame of reference is fixed and aligned with the Cartesian coordinate axes, whose origin is the center of the domain (see also \Cref{fig:model:domain:details:coordinatesystems}). We train a model which predicts the flow field at the specific instance in time when the stirrer blades are exactly aligned with the $x$ and $y$ axes. As the stirrer is assumed to rotate with a constant angular velocity $\omega$, a linear velocity profile develops on the stirrer blades according to the law of uniform circular motion. This results in the following \glspl{bc}: Due to the no-slip condition, the fluid adheres to the stirrer on $\Gamma_\text{stirrer}$, i.e., $\bm{v} = \bm{v}_\text{stirrer}$, where $\bm{v}_\text{stirrer}$ can be computed as 
\begin{align}
    \label{eq:model:domain:3Dstirrervelocity}
    \bm{v}_\text{stirrer} &= \bm{\omega} \times \bm{r} \; ,
\end{align}
where $\bm{\omega}$ is the rotational velocity vector and $\bm{r}$ the radial vector originating from the center of rotation. In 2D, only rotation around the $z$-axis is possible, i.e., $\bm{\omega}=\omega\bm{e}_z$, with $\bm{e}_z$ denoting the unit vector in $z$-direction. Consequently, \Cref{eq:model:domain:3Dstirrervelocity} simplifies to
\begin{align}
    \label{eq:model:domain:2Dstirrervelocity}
    \bm{v}_\text{stirrer} &= \begin{pmatrix} \phantom{-}\omega y \\ -\omega x \end{pmatrix} \; ,
\end{align}
if the origin of the $xy$-coordinate system is located on the stirrer axis. On the reactor wall $\Gamma_\text{wall}$, we employ the no-slip condition, i.e., $\bm{v}=\bm{0}$.

This allows us to define the following \gls{bc} residuals that are required for the loss function of the \gls{pinn} \Cref{eq:methods:pinn:physicslosses:bcs} as $\bm{\mathcal{B}}[\bm{u}]=(\bm{\mathcal{B}}_\text{wall}[\bm{u}]^T,\bm{\mathcal{B}}_\text{stirrer}[\bm{u}]^T)^T$, with
\begin{subequations}
\label{eq:model:domain:bcs}
\begin{align}
    \label{eq:model:domain:bcs:wall}
    \bm{\mathcal{B}}_\text{wall}[\bm{u}] &= \begin{pmatrix} v_x \\ v_y \end{pmatrix} \; , \\
    \label{eq:model:domain:bcs:stirrer}
    \bm{\mathcal{B}}_\text{stirrer}[\bm{u}] &= \begin{pmatrix} v_x - \omega y \\ v_y + \omega x \end{pmatrix} \; .
\end{align}
\end{subequations}

For some models presented in \Cref{sec:nonparameterized,sec:parameterized}, we will consider the problem domain in polar instead of Cartesian coordinates. Both coordinate systems are visualized in \Cref{fig:model:domain:details:coordinatesystems}. We will denote the Cartesian input vector by $\bm{x}_\text{cart}\coloneqq(x,y)^T\in\Omega_\text{cart}$ and write $\bm{x}_\text{polar}\coloneqq(r,\varphi)^T\in\Omega_\text{polar}$ in the case of polar coordinates. For the case of the parameterized model presented in \Cref{sec:parameterized}, we only consider polar coordinates but extend the polar domain $\Omega_\text{polar}$ by the parameter space $\Lambda$. Consequently, we denote the input vector as $\bm{x}_\text{param} \coloneqq (r, \varphi, \omega)^T\in\Omega_\text{polar}\times\Lambda$.
\begin{figure}[ht]
    \centering
    \begin{subfigure}[t]{0.47\textwidth}
        \centering
\setlength{\gridlength}{20cm}

\begin{tikzpicture}[font=\large, x=\gridlength, y=\gridlength]
    
    \def \sqrttwo {1.41421356237}
    \def \position {0.1/\sqrttwo}
    
    \draw[thick, draw=black] (0,0) circle (0.1);
    
    \begin{scope}
        \clip (0,0) circle (0.1);
        \foreach \x/\y/\d in {\position/\position/-45, \position/-\position/45, -\position/-\position/-45, -\position/\position/45} {
            \node[thick, draw=black, fill=black!10, shape=rectangle, minimum width=0.005\gridlength, minimum height=0.03\gridlength, inner sep=0pt, anchor=center, rotate=\d] at (\x,\y) {};
        }
    \end{scope}
    
    \foreach \r in {-0.04, 0.04} {
        \draw[thick] (0,0) -- (0,\r);
        \draw[thick] (0,0) -- (\r,0);
    }
    
    
    \draw[custom-blue-3, very thick, -{latex}] (0,0) -- (0.55*\position,0) node[below] {$x$};
    \draw[custom-blue-3, very thick, -{latex}] (0,0) -- (0,0.55*\position) node[left] {$y$};
    
    \draw[custom-red, very thick, -{latex}] (0,0) -- (0.55*\position,0.55*\position);
    \draw[custom-red, very thick, -{latex}] (0.0425,0) to [out=90,in=315, looseness=1] (0.5*\position-0.005,0.5*\position-0.005);
    \node[custom-red, above] at (0.55*\position,0.55*\position) {$r$};
    \node[custom-red, below right] at (0.55*\position,0.55*\position-0.01) {$\varphi$};
    
\end{tikzpicture}
        \caption{Definition of Cartesian and polar coordinate systems for the  problem domain.}
        \label{fig:model:domain:details:coordinatesystems}
    \end{subfigure}
    \hfill
    \begin{subfigure}[t]{0.47\textwidth}
        \centering
\setlength{\gridlength}{20cm}

\begin{tikzpicture}[font=\large, x=\gridlength, y=\gridlength]
    
    \def \sqrttwo {1.41421356237}
    \def \position {0.1/\sqrttwo}
    
    \draw[thick, draw=black] (0,0) circle (0.1);
    
    \begin{scope}
        \clip (0,0) circle (0.1);
        \foreach \x/\y/\d in {\position/\position/-45, \position/-\position/45, -\position/-\position/-45, -\position/\position/45} {
            \node[thick, draw=black, fill=black!10, shape=rectangle, minimum width=0.005\gridlength, minimum height=0.03\gridlength, inner sep=0pt, anchor=center, rotate=\d] at (\x,\y) {};
        }
    \end{scope}
    
    \foreach \r in {-0.04, 0.04} {
        \draw[thick] (0,0) -- (0,\r);
        \draw[thick] (0,0) -- (\r,0);
    }
    
    
    \draw[custom-blue-3, very thick, dashed, -{latex}|] (0,0) -- (\position-0.01,0.01-\position) node[left] {$R_\text{baffle}$};
    
    \draw[custom-blue-3, very thick, dashed, -{latex}|] (0,0) -- (0.1,0.0) node[above left, text width=20mm, align=right] {$R_\text{reactor}$};
    
    \draw[custom-blue-3, very thick, dashed, -{latex}|] (0,0.0075) -- (-0.04,0.0075) node[above] {$R_\text{stirrer}$};
    
    \coordinate (AH) at (-0.0769, -0.0433);
    \coordinate (AT) at (-0.0619, -0.0583);
    \coordinate (BH) at (-0.0433, -0.0769);
    \coordinate (BT) at (-0.0583, -0.0619);
    \draw[custom-blue-3, very thick, dashed, -{latex}|] (AH) -- (AT);
    \draw[custom-blue-3, very thick, dashed, -{latex}|] (BH) -- (BT);
    \node[custom-blue-3, right] at (0.015-\position,0.015-\position) {$t_\text{baffle}$};
    
\end{tikzpicture}
        \caption{Geometrical dimensions of the problem domain.}
        \label{fig:model:domain:details:dimensions}
    \end{subfigure}
    \caption{Coordinate systems and geometrical dimensions related to the problem domain.}
    \label{fig:model:domain:details}
\end{figure}

The geometry of our domain can be described by geometrical quantities visualized in \Cref{fig:model:domain:details:dimensions}. Their numerical values are listed in \Cref{tab:model:domain:data:dimensions}.
\begin{table}[ht]
    \centering
    \caption{Geometrical dimensions and material properties that are considered constant throughout the paper.}
    \label{tab:model:domain:data}
    \begin{subtable}[t]{0.45\textwidth}
        \centering
        \caption{Geometrical dimensions of the \gls{str} depicted in \Cref{fig:model:domain:details:dimensions}.}
        \label{tab:model:domain:data:dimensions}
        \begin{tabular}{lc}
            \toprule
            Quantity & Value $[\unit{\meter}]$ \\
            \midrule
            $R_\text{stirrer}$ & \num{0.040} \\
            $R_\text{baffle}$ & \num{0.085} \\
            $R_\text{reactor}$ & \num{0.100} \\
            $t_\text{baffle}$ & \num{0.005} \\
            \bottomrule
        \end{tabular}
    \end{subtable}
    \hfill
    \begin{subtable}[t]{0.45\textwidth}
        \centering
        \caption{Constant material properties used for the simulations and \gls{pinn} training.}
        \label{tab:model:domain:data:materialproperties}
        \begin{tabular}{lll}
            \toprule
            Property & Value & Unit \\
            \midrule
            $\rho$ & \num{1000}  & \unit{\kilogram\per\meter\cubed} \\
            $\mu$ & \num{0.001} & \unit{\kilogram\per\meter\per\second} \\
            \bottomrule
        \end{tabular}
    \end{subtable}
\end{table}
\Cref{tab:model:domain:data:materialproperties} lists the values for the density and viscosity which we consider constant throughout this work. Note that the stirrer velocity $\omega$ is not listed in either table. While we consider a fixed stirrer velocity in \Cref{sec:nonparameterized} corresponding to a Reynolds number of $\mathrm{Re}=\num{4000}$, in \Cref{sec:parameterized} we treat it as a variable parameter and train \glspl{pinn} that can predict the flow field for different impeller speeds. The Reynolds number, which describes the ratio of inertial to viscous forces in a fluid, is an important non-dimensional quantity to characterize flow regimes. In this paper, we follow the definition from \cite{Rosseburg2018}, i.e.,
\begin{align}
    \label{eq:model:domain:reynolds}
    \mathrm{Re} &= \frac{\rho\omega(2R_\text{stirrer})^2}{\mu} \; .
\end{align}
From \Cref{eq:model:domain:reynolds} we can see that the operating point of $\mathrm{Re}=\num{4000}$ corresponds to a stirrer velocity of $\omega=\qty{0.625}{\radian\per\second}$.


\subsection{Validation}
\label{subsec:model:validation}
The models trained within the scope of this paper were validated against reference solutions obtained from high-fidelity simulations using our in-house solver XNS \cite{Dirkes2024}. The problem defined above was discretized on an unstructured mesh $\Omega_\text{ref}\subset\Omega$ consisting of \num{73164} triangular elements and $N_\text{ref}=\num{37117}$ nodes and solved using a stabilized FEM approach \cite{Tezduyar1992} with linear basis functions for the pressure and velocity fields. High-fidelity solutions were obtained for \\ $\mathrm{Re} \in \{\num{1000}, \num{4000}, \num{6000}, \num{8000}, \num{10000}\}$. Reference solutions for velocity and pressure for $\mathrm{Re}=\num{4000}$ are shown in \Cref{fig:model:referencesol4000}.
\begin{figure}[ht]
    \centering
    \begin{subfigure}[t]{0.47\textwidth}
        \centering
        \input{figures/reference/reynolds_4000/vel_mag.pgf}
        \caption{Magnitude of the velocity field $\|\bm{v}\|_2$.}
        \label{fig:model:referencevel4000}
    \end{subfigure}
    \hfill
    \begin{subfigure}[t]{0.47\textwidth}
        \centering
        \input{figures/reference/reynolds_4000/p.pgf}
        \caption{Pressure field $p$.}
        \label{fig:model:referencepress4000}
    \end{subfigure}
    \caption{High-fidelity solution for velocity (A) and pressure (B) at $\mathrm{Re}=\num{4000}$ that was used to validate predictions of the \gls{pinn} models.}
    \label{fig:model:referencesol4000}
\end{figure}

We used point-wise $\ell^1$ and $\ell^2$ errors as defined in \Cref{eq:model:validation:errormetrics} as metrics to evaluate the performance of the models. Each trained \gls{pinn} model was employed to make predictions for a randomly selected subset of $N_\text{eval}=\num{10000}$ nodes $\bm{X}_\text{eval} \subset \Omega_\text{ref}$ from the reference mesh. While we can directly compute the discrepancy between the predicted velocity field and the reference velocity, the pressure predictions cannot be directly compared to the reference solution: Since our problem formulation does not constrain the pressure in the computational domain, we first need to ensure that the predicted pressure and reference values are on the same scale. To this end, we apply an affine transformation so that the pressure maximum is zero for both the \gls{pinn} predictions and the reference solution, i.e., 
\begin{subequations}
\label{eq:model:validation:pressuretransform}
\begin{align}
    \tilde{p}^\text{PINN}(\bm{x}) &= p^\text{PINN}(\bm{x}) - p^\text{PINN}_\text{max} \; , \\
    \tilde{p}^\text{ref}_{\bm{x}} &= p^\text{ref}_{\bm{x}} - p^\text{ref}_\text{max} \; ,
\end{align}
\end{subequations}
with $p^{(\cdot)}_\text{max}=\max\limits_{\bm{x}\in\bm{X}_\text{eval}}p^{(\cdot)}$.
Subsequently, we can define the point-wise error function for a quantity $q\in\{\tilde{p},\bm{v}\}$ as
\begin{align}
\label{eq:model:validation:errorfunction}
    f_\text{err}^{(q)}(\bm{x}) = ||q^\text{PINN}(\bm{x})||_2 - ||q^\text{ref}_{\bm{x}}||_2 \; .
\end{align}
\textit{Remark:} In case of a vector-valued quantity, its 2-norm is defined as $\|q\|_2\coloneqq\sqrt{q_x^2+q_y^2}$, while for a scalar quantity, the 2-norm is equivalent to the absolute value $\|q\|_2\equiv|q|$.

With that, we compute the (normalized) $\ell^1$ and $\ell^2$ errors of our models as
\begin{subequations}
\label{eq:model:validation:errormetrics}
\begin{align}
    \delta_{\ell^1}^{(q)} &= \frac{1}{q_\text{norm}}\frac{1}{N_\text{eval}} \sum_{k=1}^{N_\text{eval}}\left|\bm{f}^{(q)}_\text{err}(\bm{x}_k)\right| \; , \\
    \delta_{\ell^2}^{(q)} &= \frac{1}{q_\text{norm}}\sqrt{\frac{1}{N_\text{eval}} \sum_{k=1}^{N_\text{eval}}\left(\bm{f}^{(q)}_\text{err}(\bm{x}_k)\right)^2} \; .
\end{align}
\end{subequations}
Here, $q_\text{norm}$ stands for a normalizing quantity to which the errors are relative to. We use the velocity at the tip of the stirrer, $v_\text{norm}\equiv v_\text{tip}\coloneqq \omega R_\text{stirrer}$, and the range of reference pressure, $\tilde{p}_\text{norm}\equiv p^\text{ref}_\Delta \coloneqq p^\text{ref}_\text{max} - p^\text{ref}_\text{min}$ as normalizing quantities for $\delta^{(\bm{v})}$ and $\delta^{(\tilde{p})}$, respectively.


\section{Non-parameterized case} 
\label{sec:nonparameterized}
The initial stage of our work focused on exploring various strategies to improve the prediction quality of \glspl{pinn} for our application. The non-parameterized models were trained for fixed $\mathrm{Re}=\num{4000}$. The best performing model identified in this section was then extended through the introduction of the rotational velocity of the stirrer as a parametric input. This section presents a series of \gls{pinn} configurations, each increasing in complexity as we progressively incorporate insights into the physics of the problem into the model. Supplementary information for all models which is not explicitly commented on in the following sections can be found in the model-respective section in \Cref{app:nonparameterized}.


\subsection{Baseline model}
\label{subsec:nonparameterized:baseline}
In the most basic \gls{pinn} configuration, the model is trained on the domain $\Omega_\text{cart}$, with the input vector to the \gls{nn}, $\bm{x}_\text{cart}$, consisting of Cartesian coordinate pairs for the training points. The loss function comprises the residuals of the conservation equations as defined in \Cref{eq:model:governingequations} and \gls{bc} residuals for $\Gamma_\text{wall}$ and $\Gamma_\text{stirrer}$ as defined in \Cref{eq:model:domain:bcs}. Scaling factors for all loss components are set to one. The pre-processing function $\bm{f}_\text{pre}$ from \Cref{eq:methods:inputoutputscaling:ansatz} is chosen as
\begin{align}
    \label{eq:nonparameterized:baseline:preprocessing}
    \tilde{\bm{x}} &= \bm{f}_\text{pre}(\bm{x}_\text{cart}) = \begin{pmatrix} \nicefrac{x}{R_\text{reactor}} \\ \nicefrac{y}{R_\text{reactor}} \end{pmatrix} \; .
\end{align}
The non-dimensional outputs of the network are scaled as follows:
\begin{align}
    \label{eq:nonparameterized:baseline:postprocessing}
    \bm{u} &= \bm{f}_\text{post}(\tilde{\bm{u}}) = \begin{pmatrix} \phantom{\rho}v_\text{tip} \tilde{\bm{v}} \\ \rho v_\text{tip}^2 \tilde{p} \end{pmatrix} \; .
\end{align}
\Cref{fig:nonparameterized:baseline:err} depicts the spatial distribution of the normalized velocity (\Cref{fig:nonparameterized:baseline:err:vmagrel4000}) and pressure errors (\Cref{fig:nonparameterized:baseline:err:prel4000}). 
\begin{figure}[ht]
    \centering
    \begin{subfigure}[t]{0.47\textwidth}
        \centering
        \input{figures/01_vanilla/reynolds_4000/err_vmag_rel_contour.pgf}
        \caption{Normalized error in velocity magnitude.}
        \label{fig:nonparameterized:baseline:err:vmagrel4000}
    \end{subfigure}
    \hfill
    \begin{subfigure}[t]{0.47\textwidth}
        \centering
        \input{figures/01_vanilla/reynolds_4000/err_p_rel_contour.pgf}
        \caption{Normalized error in pressure.}
        \label{fig:nonparameterized:baseline:err:prel4000}
    \end{subfigure}
    \caption{Spatial distribution of normalized errors of the baseline model presented in \Cref{subsec:nonparameterized:baseline} for the velocity (A) and pressure fields (B). The velocity errors are most pronounced in the area between the stirrer and the baffles, where the \gls{pinn} solely relies on the \gls{pde} residuals without information from the labeled boundary points, while the highest pressure errors are localized in the inner region around the stirrer.}
    \label{fig:nonparameterized:baseline:err}
\end{figure}
It is evident from \Cref{fig:nonparameterized:baseline:err} that the baseline model struggles to accurately capture both the velocity and pressure fields, with the maximum normalized velocity error reaching approximately \qty{36.2}{\percent}. The pressure errors are most pronounced in the inner region around the stirrer ($r \lesssim R_\text{stirrer}$), while the errors in velocity magnitude are highest in the area between the stirrer and the baffles. In this region, the \gls{pinn} relies solely on the \gls{pde} residuals without information from the labeled points on the boundary. In the further refinements of the model, we address this issue initially by incorporating labeled training data for the domain in \Cref{subsec:nonparameterized:baselinedata} and subsequently by adjusting the scaling of the loss components corresponding to the \gls{pde} residuals in \Cref{subsec:nonparameterized:baselinescaling}. Moreover, we observe that despite the symmetry of the problem domain, the learned solution lacks this property. We will address this shortcoming with a more advanced model in \Cref{subsec:nonparameterized:baselinepolar}. As the accuracy in capturing the velocity field holds greater significance for our application than the pressure prediction, our focus will be on analyzing the velocity error and we will utilize it as the primary measure of model quality in the remainder of this paper.


\subsection{Baseline model with data}
\label{subsec:nonparameterized:baselinedata}
Harnessing the ability of \glspl{pinn} to integrate both physics and data losses (cf. \Cref{eq:methods:pinn:physicsdataloss}), we introduce \num{2000} labeled training points obtained from the reference solution described in \Cref{subsec:model:validation}. \Cref{fig:nonparameterized:baselinedata:err:vmagrel4000} illustrates that incorporating this additional information into the training notably enhances the predictive performance of the model detailed in \Cref{subsec:nonparameterized:baseline}. This improvement is evident in the reduction of both the maximum and average velocity error values, particularly in the domain far away from the boundaries. Nevertheless, it is reasonable to assume that acquiring such a substantial amount of data for each distinct set of process parameters would be prohibitively costly. Therefore, throughout the remainder of this section, we will focus on investigating strategies aimed at improving the performance of the model without the need for labeled data.
\begin{figure}[ht]
    \centering
    \begin{subfigure}[t]{0.47\textwidth}
        \centering
        \input{figures/02_vanilla-data/reynolds_4000/err_vmag_rel_contour.pgf}
        \caption{Normalized error in velocity magnitude of vanilla model with labeled data.}
        \label{fig:nonparameterized:baselinedata:err:vmagrel4000}
    \end{subfigure}
    \hfill
    \begin{subfigure}[t]{0.47\textwidth}
        \centering
        \input{figures/03_vanilla-scaled/reynolds_4000/err_vmag_rel_contour.pgf}
        \caption{Normalized error in velocity magnitude of baseline model with scaled loss components.}
        \label{fig:nonparameterized:baselinescaled:err:vmagrel4000}
    \end{subfigure}
    \caption{Spatial distribution of normalized velocity errors of the baseline model with labeled training data (A) and of the baseline model with scaled loss components (B).}
    \label{fig:nonparameterized:baselinescaled:err}
\end{figure}


\subsection{Baseline model with loss scaling}
\label{subsec:nonparameterized:baselinescaling}
As mentioned in \Cref{subsec:methods:pinn}, the prediction capabilities of a \gls{pinn} can be significantly improved by scaling the loss components appropriately. In this work, the scaling factors were manually tuned based on insights gained from the reference solution. As depicted in \Cref{fig:nonparameterized:baseline:err:vmagrel4000}, utilizing unit scaling for the loss components results in a relatively low velocity error near $\Gamma_\text{stirrer}$, a slightly larger error near $\Gamma_\text{wall}$, while the most significant errors are observed further away from the boundaries. Based on this observation, we increase the scaling factors of $\bm{\alpha}_\text{PDE}=(\alpha_{\text{momentum}_x}, \alpha_{\text{momentum}_y}, \alpha_\text{mass})^T$ to $(50, 50, 50)^T$ and set $\alpha_\text{wall} \coloneqq 5$. \Cref{fig:nonparameterized:baselinescaled:err:vmagrel4000} shows that just by scaling the loss components we were able to significantly improve the accuracy of the velocity predicted by the \gls{pinn}. The error for predictions within the domain between the stirrer tip and the wall now ranges from \qtyrange{5}{10}{\percent}, compared to over \qty{30}{\percent} observed with unit scaling in \Cref{subsec:nonparameterized:baseline}. The highest error is currently localized around the stirrer, as the loss component corresponding to the stirrer \gls{bc} carries the least weight.


\subsection{Baseline model in polar coordinates}
\label{subsec:nonparameterized:baselinepolar}
Due to the geometry of the cross-section depicted in \Cref{fig:model:domain:overview:2d}, polar coordinates are the more intuitive coordinate system for the current application. Consequently, instead of solving for velocities in $x$- and $y$-direction $\bm{v}_\text{cart}=(v_x,v_y)^T$, we solve for the radial and circumferential velocity components $\bm{v}_\text{polar}=(v_r,v_\varphi)^T$. We can also leverage the fact that the geometry is symmetric to reduce the computational domain to $\Omega_\text{sym}\subset\Omega_\text{polar}$ depicted in \Cref{fig:nonparameterized:dd:overview:domains}, which constitutes a quarter of the original domain. The bounds of $\Omega_\text{sym}$ are defined by the axes connecting the opposing baffles. The input coordinates $\bm{x}_\text{polar}$ are then normalized such that
\begin{align}
\label{eq:nonparameterized:baselinepolar:preprocessing}
    \tilde{\bm{x}} &= \bm{f}_\text{pre}(\bm{x}_\text{polar}) = \begin{pmatrix} \nicefrac{(r-R_\text{stirrer})}{(R_\text{reactor}-R_\text{stirrer})} \\ \nicefrac{4\varphi}{\pi} \end{pmatrix} \; .
\end{align} 
The respective outputs of the \gls{nn} are scaled as
\begin{align}
    \bm{u} &= \bm{f}_\text{post}(\tilde{\bm{u}}) = \begin{pmatrix} v^\text{ref}_{r,\text{max}}\tilde{v}_r \\ v_\text{tip}\tilde{v}_\varphi \\ \rho v_\text{tip}^2\tilde{p} \end{pmatrix} \; ,
\end{align}
where $v^\text{ref}_{r,\text{max}}=\qty{8e-4}{\meter\per\second}$ was obtained from the high-fidelity solution on our reference mesh.
The solution in the full domain can then be inferred from the solution in $\Omega_\text{sym}$ through projection onto the remaining part of the domain. We define the additional \gls{bc} to ensure continuity of the solution on $\Gamma_\text{sym}$ as:
\begin{align}
    \label{eq:nonparameterized:polar:symmetrybc}
    \bm{u}\rvert_{\bm{x}^{(+)}} = \bm{u}\rvert_{\bm{x}^{(-)}} \quad \text{on} \; \Gamma_\text{sym} \; .
\end{align}
Here, $\bm{x}^{(+)}$ and $\bm{x}^{(-)}$ represent the left and right limiting values of spatial coordinates approaching $\Gamma_\text{sym}$.

\begin{figure}[ht]
    \centering
        \input{figures/05_vanilla-polar-scaled/reynolds_4000/err_vmag_rel_contour.pgf}
        \caption{Spatial distribution of normalized error in velocity magnitude of the vanilla model trained on $\Omega_\text{sym}$ using polar coordinates.}
        \label{fig:nonparameterized:baselinepolar:err:vmagrel4000}
\end{figure}
\Cref{fig:nonparameterized:baselinepolar:err:vmagrel4000} demonstrates that in terms of average velocity prediction error, the model at hand performs similarly to the model presented in \Cref{subsec:nonparameterized:baselinescaling}. Consequently, we can conclude that the choice of the coordinate system does not have a significant impact on the predictive capabilities. It is also evident from \Cref{fig:nonparameterized:baselinepolar:err:vmagrel4000} that the symmetry boundary constraint is not entirely satisfied. This aspect could be further improved by fine-tuning the respective loss scaling factors.

\begin{figure}[bt]
    \centering
    \begin{subfigure}[t]{0.47\textwidth}
        \centering

\setlength{\gridlength}{20cm}

\begin{tikzpicture}[font=\large, x=\gridlength, y=\gridlength]
    
    \clip (-0.105,-0.105) rectangle (0.105,0.105);
    
    \def \xThetaP {0.073612159}
    \def \yThetaP {0.0425}
    \def \sqrttwo {1.41421356237}
    \def \position {0.1/\sqrttwo}
    \def \radImp {0.04}
    \def \radIntf {0.06}
    \def \intf{ \radIntf/\sqrttwo }
    \def \baffOuter {\position}
    \def \baffInner {0.06}
    
    \fill[very thick, draw=black, fill=custom-blue-1] (0,0) circle (0.1);
    \fill[fill=custom-yellow!40] (0,0) circle (\radIntf);
    
    \draw[very thick, black] (0,0) circle (0.1);
    
    \draw[very thick, custom-red, dashed, pattern=north east lines, pattern color=custom-red, opacity=0.4] (0,0) -- (\baffOuter,-\baffOuter) arc[start angle=-45, end angle=45, radius=0.1] -- cycle;
    
    \begin{scope}
        \clip (0,0) circle (0.1);
        \def \sqrttwo {1.41421356237}
        \def \position {0.1/\sqrttwo}
        \foreach \x/\y/\d in {\position/\position/-45, \position/-\position/45, -\position/-\position/-45, -\position/\position/45} {
            \node[very thick, draw=black, fill=black!10, shape=rectangle, minimum width=0.005\gridlength, minimum height=0.03\gridlength, inner sep=0pt, anchor=center, rotate=\d] at (\x,\y) {};
        }
    \end{scope}
    
    \foreach \r in {-0.04, 0.04} {
        \draw[very thick] (0,0) -- (0,\r);
        \draw[very thick] (0,0) -- (\r,0);
    }
    
    \draw[custom-blue-3, very thick, dashed, -{latex}|] (0,0) -- (-\intf,-\intf) node[above left] {$R_\text{inter}$};
    
    \node[custom-blue-3] at (0.0,0.08) {$\Omega_\text{outer}$};
    \node[custom-orange] at (-0.025,0.025) {$\Omega_\text{inner}$};
    \node[custom-red] at (0.065,0.0) {$\Omega_\text{sym}$};
    
\end{tikzpicture}
        \caption{Division of the computational domain into an inner $\Omega_\text{inner}\subset\Omega_\text{polar}$ and outer region $\Omega_\text{outer}\subset\Omega_\text{polar}$. The interface between these two domains is located at $r=R_\text{inter}$. If a model exploits symmetry, only a quarter $\Omega_\text{sym}\subset\Omega_\text{polar}$ of the full domain $\Omega_\text{polar}$ is considered.}
        \label{fig:nonparameterized:dd:overview:domains}
    \end{subfigure}
    \hfill
    \begin{subfigure}[t]{0.47\textwidth}
        \centering

\setlength{\gridlength}{20cm}

\begin{tikzpicture}[font=\large, x=\gridlength, y=\gridlength]
    
    \clip (-0.105,-0.105) rectangle (0.105,0.105);
    
    \def \xThetaP {0.073612159}
    \def \yThetaP {0.0425}
    \def \sqrttwo {1.41421356237}
    \def \position {0.1/\sqrttwo}
    \def \radImp {0.04}
    \def \radIntf {0.06}
    \def \intf {\radIntf/\sqrttwo}
    \def \baffOuter {\position}
    \def \baffInner {0.06}
    
    \fill[very thick, draw=black, fill=custom-blue-1] (0,0) circle (0.1);
    
    \fill[fill=white, opacity=0.6] (0,0) -- (\baffOuter,\baffOuter) arc[start angle=45, end angle=315, radius=0.1] -- cycle;
    
    \fill[fill=custom-yellow!40] (0,0) circle (\radIntf);
    
    \draw[very thick, black] (0,0) circle (0.1);
    
    \draw[very thick, custom-orange, dashed] (0,0) circle (\radIntf);
    \draw[very thick, custom-orange] (\intf,-\intf) arc[start angle=-45, end angle=45, radius=\radIntf];
    \draw[thick, custom-blue-2] (\baffInner,-\baffInner) arc[start angle=-45, end angle=45, radius=0.085];
    \draw[very thick, custom-blue-3, dashed, {latex}-{latex}] (\xThetaP,-\yThetaP) arc[start angle=-30, end angle=30, radius=0.085];
    
    \draw[very thick, custom-red, dashed] (\intf, \intf) -- (\baffInner,\baffInner);
    \draw[very thick, custom-red, dashed] (\intf,-\intf) -- (\baffInner,-\baffInner);
    
    \begin{scope}
        \clip (0,0) circle (0.1);
        \def \sqrttwo {1.41421356237}
        \def \position {0.1/\sqrttwo}
        \foreach \x/\y/\d in {\position/\position/-45, \position/-\position/45, -\position/-\position/-45, -\position/\position/45} {
            \node[very thick, draw=black, fill=black!10, shape=rectangle, minimum width=0.005\gridlength, minimum height=0.03\gridlength, inner sep=0pt, anchor=center, rotate=\d] at (\x,\y) {};
        }
    \end{scope}
    
    \foreach \r in {-0.04, 0.04} {
        \draw[very thick] (0,0) -- (0,\r);
        \draw[very thick] (0,0) -- (\r,0);
    }
    
    \node[custom-orange] at (0.032,0.014) {$\Gamma_\text{inter}$};
    \node[custom-red] at (0.034,-0.054) {$\Gamma_\text{sym}$};
    \node[custom-blue-2] at (0.062,0.038) {$\Gamma_\text{c}$};
    \node[custom-blue-3] at (0.073,0.0) {$\Gamma_\text{d}$};
    
\end{tikzpicture}
        \caption{Relevant interfaces used in the \acrshort{dd} models. $\Gamma_\text{inter}$ represents the interface between the inner and the outer domain, on $\Gamma_\text{c}$ only the continuity of values is imposed while $\Gamma_\text{d}$ additionally considers the continuity of derivatives. $\Gamma_\text{sym}$ corresponds to the interface of the symmetry domain $\Omega_\text{sym}$.}
        \label{fig:nonparameterized:dd:overview:interfaces}
    \end{subfigure}
    \caption{Schematic representation of the different subdomains and interfaces used in the more complex \gls{pinn} configurations.}
    \label{fig:nonparameterized:dd:overview}
\end{figure}


\subsection{Strong \gls{bc} model}
\label{subsec:nonparameterized:hardbc}
In all the models presented thus far, the most significant errors are localized near the tip of the stirrer and in its close proximity. A closer inspection of the velocity profiles along the radius of the domain for $\varphi=0$, compared to the reference solution in \Cref{fig:nonparameterized:baseline:velprofiles}, indicates that the error stems from the inability of the \gls{pinn} to capture the sharp kink at the stirrer's tip $r=R_\text{stirrer}$. Consequently, the Dirichlet \gls{bc} at $\Gamma_\text{stirrer}$ is violated. 
\begin{figure}[ht]
    \input{figures/90_velocity-profiles/vel_profiles_vanilla_phi0.pgf}
    \caption{Velocity magnitudes of baseline models in radial direction for $\varphi=0$. These models fail to capture the kink in velocity magnitude at the tip of the stirrer. This shortcoming is addressed by enforcing the Dirichlet \gls{bc} at $\Gamma_\text{stirrer}$ in a strong manner in \Cref{subsec:nonparameterized:hardbc,subsec:nonparameterized:hybridbc}.}
    \label{fig:nonparameterized:baseline:velprofiles}
\end{figure}
To address this issue and to reproduce the kink more accurately, we explore the application of strong \glspl{bc} as defined in \Cref{subsec:methods:strongDirichlet}. To impose the \glspl{bc} for the velocity defined in \Cref{eq:model:domain:bcs} in a strong manner through \Cref{eq:methods:pinn:approximation:strong}, we construct the function $\bm{g}_\text{strong}$ using the approach outlined in \cite{Sheng2021LFF} such that
\begin{align}
    \begin{cases}
        \bm{g}_\text{strong}(\bm{x}) = 0 & \text{on} \; \Gamma_\text{stirrer} \cup \Gamma_\text{wall} \\
        0 < \bm{g}_\text{strong}(\bm{x}) \leq 1 & \text{in} \; \Omega \setminus (\Gamma_\text{stirrer} \cup \Gamma_\text{wall}) \; .
    \end{cases}
\end{align}
Since the Dirichlet \glspl{bc} only apply to velocity, the boundary value function from \Cref{eq:methods:pinn:approximation:strong} is defined as $\bm{h}(\bm{x})\coloneqq \bm{v}_\text{BC}$. Contrary to the findings reported in \cite{Sheng2021LFF}, our research indicates that the predictive capability of the model is influenced by the choice of $\bm{h}$. Hence, we attempt to employ a $\bm{v}_\text{BC}$ that approximates the physical behavior to a certain extent. The selected function is illustrated in \Cref{fig:nonparameterized:hardbc:overview:vbc} and its derivation is provided in \Cref{app:vbc}. In this model, the input normalization and output scaling outlined in \Cref{subsec:nonparameterized:baseline} were employed. The post-processing function defined by \Cref{eq:methods:pinn:approximation:strong} was only applied to the velocity predictions after the outputs of the \gls{nn} had been rescaled.
\begin{figure}[ht]
    \centering
    \begin{subfigure}[t]{0.47\textwidth}
        \centering
        \input{figures/06x_hard-BC/reynolds_4000/vel_bc_lff.pgf}
        \caption{Magnitude of the analytical function $\bm{v}_\text{BC}$ used as $\bm{h}$ in \Cref{eq:methods:pinn:approximation:strong} for strong imposition of \glspl{bc}.}
        \label{fig:nonparameterized:hardbc:overview:vbc}
    \end{subfigure}
    \hfill
    \begin{subfigure}[t]{0.47\textwidth}
        \centering
        \input{figures/06x_hard-BC/reynolds_4000/err_vmag_rel_contour.pgf}
        \caption{Spatial distribution of normalized error in velocity magnitude for the model using strong imposition of Dirichlet \glspl{bc} presented in \Cref{subsec:nonparameterized:hardbc}. The error in velocity prediction around the stirrer is reduced, while the error near the wall and the baffles is high.}
        \label{fig:nonparameterized:hardbc:overview:err:vmagrel4000}
    \end{subfigure}
    \caption{Analytical function satisfying the Dirichlet \glspl{bc} in the domain (A) and spatial distribution of the normalized error of the model with strongly enforced \glspl{bc} (B).}
    \label{fig:nonparameterized:hardbc:overview}
\end{figure}
We can observe from \Cref{fig:nonparameterized:hardbc:overview:err:vmagrel4000} that the strong imposition of the \gls{bc} on $\Gamma_\text{stirrer}$ significantly aids in reducing the velocity prediction error around the stirrer. However, the error near $\Gamma_\text{wall}$ and at the baffles is high. This could be attributed to the underlying boundary function $\bm{v}_\text{BC}$ not accurately representing the solution in this region. In the following section, we tackle this issue by maintaining strong \gls{bc} imposition for the stirrer while reverting to soft \glspl{bc} for the wall.


\subsection{Hybrid \gls{bc} model}
\label{subsec:nonparameterized:hybridbc}
To achieve better prediction capabilities and mitigate high errors in the region adjacent to the reactor wall, we choose to impose the \gls{bc} on $\Gamma_\text{stirrer}$ in a strong manner and utilize a soft \gls{bc} imposition on $\Gamma_\text{wall}$. For this purpose, we define $\bm{g}_\text{hybrid}$ such that
\begin{align}
    \begin{cases}
        \bm{g}_\text{hybrid}(\bm{x}) = 0 & \text{on} \; \Gamma_\text{stirrer}  \\
        0 < \bm{g}_\text{hybrid}(\bm{x}) \leq 1 & \text{in} \; \Omega \setminus \Gamma_\text{stirrer}
    \end{cases} \; .
\end{align}
As demonstrated in \Cref{fig:nonparameterized:hybridbc:err:velmagrel4000}, this approach effectively reduces the overall error compared to the models presented thus far. \Cref{fig:nonparameterized:baseline:velprofiles} illustrates that imposing strong \gls{bc} on $\Gamma_\text{stirrer}$ helps to capture the kink in the velocity profile accurately. The strong imposition of \glspl{bc} also reduces the number of loss components that need scaling, consequently decreasing the number of hyperparameters requiring optimization. It is worth noting, however, that this comes at the expense of a significantly increased training time, as we will later discuss in \Cref{tab:nonparameterized:summary:models} in \Cref{subsec:nonparameterized:summary}.
\begin{figure}[ht]
    \centering
    \input{figures/07_hybrid-BC/reynolds_4000/err_vmag_rel_contour.pgf}
    \caption{Spatial distribution of normalized error in velocity magnitude predicted by the model using hybrid imposition of Dirichlet \glspl{bc} described in \Cref{subsec:nonparameterized:hybridbc}.}
    \label{fig:nonparameterized:hybridbc:err:velmagrel4000}
\end{figure}


\subsection{\Acrfull{dd} model}
\label{subsec:nonparameterized:domaindecomposed2}
To further refine the model and improve the quality of the predictions, we can integrate additional insights from the reference solution. Notably, we observe minimal changes in both velocity and pressure in the circumferential direction within an inner region around the stirrer, denoted by $\Omega_\text{inner}$ in \Cref{fig:nonparameterized:dd:overview:domains}. Additionally, the magnitude of the circumferential velocity significantly exceeds that of the radial velocity component, i.e.,
\begin{align}
    \label{eq:nonparameterized:dd:assumptions}
    \frac{\partial(\,\cdot\,)}{\partial \varphi} = 0 \quad \wedge \quad v_{\varphi} \gg v_r \quad \text{in} \; \Omega_\text{inner} \; .
\end{align}
The governing equations as defined in \Cref{eq:model:governingequations} and expressed in polar coordinates, can consequently be simplified into the following system of \glspl{ode}:
\begin{subequations}
\label{eq:nonparameterized:dd:ode}
\begin{align}
    \label{eq:nonparameterized:dd:ode:p}
    \frac{v_\varphi^2}{r} - \frac{1}{\rho}\frac{\mathrm{d}p}{\mathrm{d}r} &=0 \quad \text{in} \; \Omega_\text{inner} \; ,\\
    \label{eq:nonparameterized:dd:ode:vphi}
    \frac{\mathrm{d}^2 v_\varphi}{\mathrm{d} r^2} + \frac{1}{r}\frac{\mathrm{d} v_\varphi}{\mathrm{d}r} - \frac{v_\varphi}{r^2} &= 0 \quad \text{in} \; \Omega_\text{inner} \; .
\end{align}
\end{subequations}
However, the assumptions defined in \Cref{eq:nonparameterized:dd:assumptions} are not valid in $\Omega_\text{outer}$; there, \Cref{eq:model:governingequations} must be solved. This insight motivates the idea of employing a \gls{dd} approach, adapting the \gls{xpinn} approach as outlined in \Cref{subsec:methods:advanced}. The decomposition is depicted in \Cref{fig:nonparameterized:dd:overview:domains}. We simultaneously train two \glspl{nn}, solving only \Cref{eq:nonparameterized:dd:ode} in $\Omega_\text{inner}$ to obtain solution variables $\bm{u}_\text{inner}=(v_{\text{inner},\varphi}, p_\text{inner})^T$ and solving \Cref{eq:model:governingequations} in $\Omega_\text{outer}$ to obtain $\bm{u}_\text{outer}=(v_{\text{outer},r}, v_{\text{outer},\varphi}, p_\text{outer})^T$. The location of the interface between the two domains $R_\text{inter}$ is a hyperparameter. For the model presented in this section we set $R_\text{inter}=\qty{0.07}{\meter}$. We employ strong Dirichlet \gls{bc} imposition on $\Gamma_\text{stirrer}$ as defined in \Cref{subsec:nonparameterized:hybridbc}. The function used to normalize the inputs and rescale the outputs can be found in \Cref{app:nonparameterized:domaindecomposed2}. To ensure continuity of the solution variables and their derivatives, the following constraints are imposed on $\Gamma_\text{inter}=\partial \Omega_\text{inner}$:
\begin{subequations}
\label{eq:nonparameterized:dd:interfacebc}
\begin{align}
    v_{\text{outer},r} &= 0 \quad &\text{on} \; \Gamma_\text{inter} \; , 
    \label{eq:nonparameterized:dd:interfacebc:vr}\\
    v_{\text{outer},\varphi} &= v_{\text{inner},\varphi} \quad &\text{on} \; \Gamma_\text{inter} \; , \\
    p_\text{outer} &= p_\text{inner} \quad &\text{on} \; \Gamma_\text{inter} \; , \\
    \frac{\partial v_{\text{outer},\varphi}}{\partial r} &= \frac{\mathrm{d} v_{\text{inner},\varphi}}{\mathrm{d}r} \quad &\text{on} \; \Gamma_\text{inter} \; .
\end{align}
\end{subequations}
Since we solve a different set of equations in each subdomain, we omit the requirement of residual equality on the interface. As depicted in \Cref{fig:nonparameterized:dderr:dd2:vmagrel4000}, this approach helps to reduce the average normalized error of the velocity predictions to approximately \qty{1.5}{\percent}. Moreover, we can observe that the error distribution is such that notable errors are confined to areas near the stirrer tips and the baffles. This error is associated with the inability of the \gls{pinn} model to accurately replicate the kink in the velocity profile in radial direction at $\varphi=\nicefrac{\pi}{4}$, as illustrated in \Cref{fig:nonparameterized:ddprofile:baffle}. We tackle this source of error in the subsequent section.

\begin{figure}[ht]
    \centering
    \label{fig:nonparameterized:dderr}
    \begin{subfigure}[t]{0.47\textwidth}
        \centering
        \input{figures/08_DD-polar-2/reynolds_4000/err_vmag_rel_contour.pgf}
        \caption{Spatial distribution of the normalized error in velocity magnitude predicted by the model using \gls{dd} into two domains described in \Cref{subsec:nonparameterized:domaindecomposed2}.}
        \label{fig:nonparameterized:dderr:dd2:vmagrel4000}
    \end{subfigure}
    \hfill
    \begin{subfigure}[t]{0.47\textwidth}
        \centering
        \input{figures/09_DD-polar-3/reynolds_4000/err_vmag_rel_contour.pgf}
        \caption{Spatial distribution of the normalized error in velocity magnitude predicted by the model using \gls{dd} with a split of $\Omega_\text{outer}$ presented in \Cref{subsec:nonparameterized:domaindecomposed3}.}
        \label{fig:nonparameterized:dderr:dd3:vmagrel4000}
    \end{subfigure}
    \caption{Spatial distribution of the normalized error in velocity magnitude predicted by the models using \gls{dd}.}
\end{figure}


\subsection{\gls{dd} model with outer split}
\label{subsec:nonparameterized:domaindecomposed3}
To capture the behavior illustrated in \Cref{fig:nonparameterized:ddprofile:baffle}, we can enhance the \gls{dd} approach by further splitting $\Omega_\text{outer}$ into two parts as follows:
\begin{subequations}
\begin{align}
    & \Omega_\text{outer}^{(1)} = \{(r, \varphi) \in \Omega_\text{outer} : r \leq R_\text{baff}\} \; , \\
    \text{and} \quad & \Omega_\text{outer}^{(2)} = \{(r, \varphi) \in \Omega_\text{outer} : r > R_\text{baff}\} \; .
\end{align}
\end{subequations}
We split the circumferential velocity component by defining $v_{\text{outer},\varphi}^{(1)}$ and $v_{\text{outer},\varphi}^{(2)}$, respectively. We establish two distinct interfaces between the two regions: the continuity interface $\Gamma_\text{c}$ and the derivative interface $\Gamma_\text{d}$. On $\Gamma_\text{c}$, only the continuity constraint
\begin{align}
    v_{\text{outer},\varphi}^{(1)} &= v_{\text{outer},\varphi}^{(2)} \quad \text{on} \; \Gamma_\text{c} \; ,
\end{align}
is considered. On the derivative interface $\Gamma_\text{d}$, the continuity of normal derivatives
\begin{align}
    \frac{\partial v_{\text{outer},\varphi}^{(1)}}{\partial r} &= \frac{\partial v_{\text{outer},\varphi}^{(2)}}{\partial r} \quad \text{on} \; \Gamma_\text{d} \; ,
\end{align}
is also imposed. The outer domain split is located at $R_\text{outer} = \qty{0.0851}{\meter}$, right after the baffle. The derivative interface extends between $-0.3 \leq \varphi \leq 0.3$. As illustrated in \Cref{fig:nonparameterized:dd:overview:interfaces}, the derivative interface does not extend all the way to the baffles, thereby allowing for the accurate reproduction of the kink. Given that we define the circumferential velocity as the additional output $v_{\text{outer},\varphi}^{(2)}$ in $\Omega_\text{outer}^{(2)}$, the no-slip \gls{bc} is expressed as 
\begin{align}
    \label{eq:nonparameterized:dd3:wallbc}
    v_{\text{outer},\varphi}^{(2)} = 0 \quad \text{on} \; \Gamma_\text{wall} \; .
\end{align}
For the wall segment at the baffle, as illustrated in \Cref{fig:nonparameterized:dd3:bafflecloseup}, the no-slip \gls{bc} applies to $v_{\text{outer},\varphi}^{(1)}$, making 
\begin{align}
    \label{eq:nonparameterized:dd3:bafflebc}
    v_{\text{outer},\varphi}^{(1)} = 0 \quad \text{on} \; \Gamma_\text{baffle} \; .
\end{align}
\begin{figure}[ht]
    \centering
\setlength{\gridlength}{40cm}

\begin{tikzpicture}[font=\large, x=\gridlength, y=\gridlength]

    \def \sqrttwo {1.41421356237}
    \def \radReac {0.1}
    \def \radInt {0.0635}
    \def \baffleCenter {\radReac/\sqrttwo}
    \def \baffleTip {0.035}


    \draw[custom-blue-1, pattern={Lines[angle=45,distance={6pt}]}, pattern color=custom-blue-2] (0,0) -- (\radInt,0) to [out=125,in=325, looseness=1] (0,\radInt) -- cycle;
    \draw[custom-blue-1, pattern={Lines[angle=-45,distance={6pt}]}, pattern color=custom-blue-2] (\radInt,0) --(\radReac,0) to [out=120,in=330, looseness=1] (0,\radReac) -- (0,\radInt) to [out=325,in=125, looseness=1] (\radInt,0);
    \draw[very thick, black] (\radReac,0) to [out=120,in=330, looseness=1] (0,\radReac);

    \draw[very thick, custom-blue-2] (\radInt,0) to [out=125,in=325, looseness=1] (0,\radInt);
    
    \begin{scope}
        \clip (0,0) -- (\radReac,0) to [out=120,in=330, looseness=1] (0,\radReac) -- cycle;
        \node[very thick, draw=black, fill=black!10, shape=rectangle, minimum width=0.018\gridlength, minimum height=0.1\gridlength, inner sep=0pt, anchor=center, rotate=-45] at (\baffleCenter,\baffleCenter) {};
    \end{scope}

    \draw[line width=1mm, custom-red] (\baffleTip-0.007,\baffleTip+0.007) -- (\baffleTip+0.007,\baffleTip-0.007);


    \node[custom-blue-3] at (0.015,0.01) {$\Omega^{(1)}_\text{outer}$};
    \node[custom-blue-2] at (0.045,0.01) {$\Gamma_\text{c}$};
    \node[custom-blue-3] at (0.08,0.01) {$\Omega^{(2)}_\text{outer}$};
    \node[custom-red] at (0.025,0.03) {$\Gamma_\text{baffle}$};
    \node[black] at (0.09,0.045) {$\Gamma_\text{wall}$};
    
\end{tikzpicture}
    \caption{Close up of the upper right baffle for the \gls{dd} models. The segment of the boundary highlighted in red, $\Gamma_\text{baffle}$, uniquely belongs to $\Omega^{(1)}_\text{outer}$, i.e., we need to prescribe a \gls{bc} for $v^{(1)}_{\text{outer},\varphi}$, while the rest of the wall belongs to $\Omega^{(2)}_\text{outer}$ and we prescribe $v^{(2)}_{\text{outer},\varphi}$.}
    \label{fig:nonparameterized:dd3:bafflecloseup}
\end{figure}

Splitting the outer domain into two regions not only reduces the mean normalized error of the velocity predicted by the \gls{pinn}, but it also enhances the quality of the predicted velocity field. This improvement is illustrated in figure \Cref{fig:nonparameterized:dderr:dd3:vmagrel4000}, where there are no prominent areas with concentrated higher error levels. Moreover, \Cref{fig:nonparameterized:ddprofile} demonstrates that this model successfully replicates the velocity profile along the radius both for $\varphi=0$ (\Cref{fig:nonparameterized:ddprofile:inner}) and $\varphi=\nicefrac{\pi}{4}$ (\Cref{fig:nonparameterized:ddprofile:baffle}).
\begin{figure}[ht]
    \centering
    \begin{subfigure}[t]{0.47\textwidth}
        \centering
        \input{figures/90_velocity-profiles/vel_profiles_dd_phi0.pgf}
        \caption{Velocity profiles predicted by \gls{dd} models in radial direction for an angle of $\varphi=0$.}
        \label{fig:nonparameterized:ddprofile:inner}
    \end{subfigure}
    \hfill
    \begin{subfigure}[t]{0.47\textwidth}
        \centering
        \input{figures/90_velocity-profiles/vel_profiles_dd_phi45.pgf}
        \caption{Close-up of the kink in velocity profiles in radial direction at the baffle predicted by \gls{dd} models for an angle of $\varphi=\nicefrac{\pi}{4}$.}
        \label{fig:nonparameterized:ddprofile:baffle}
    \end{subfigure}
    \caption{Profiles of velocity magnitude in radial direction $r$ predicted by the \gls{dd} models.}
    \label{fig:nonparameterized:ddprofile}
\end{figure}


\subsection{Robustness study}
\label{subsec:nonparameterized:robustnessstudy}
Given that the training of a \gls{pinn} involves an optimization process with a highly complex (and non-convex) objective function landscape, and stagnation in saddle points is a common challenge in high-dimensional optimization problems \cite{Dauphin2014}, we had to make sure our models are robust and reproducible with the specified hyperparameter settings. To address this, we conducted a robustness study by training the same model 10 times with different random initializations of the \glspl{nn}' weights. As the model presented in \Cref{subsec:nonparameterized:domaindecomposed2} is the one chosen for subsequent parameterization, it is also employed to demonstrate the robustness study. We utilize the normalized velocity error defined in \Cref{eq:model:validation:errorfunction} along the radius of the impeller for $\varphi=0$ to showcase the model's robustness in \Cref{fig:nonparameterized:robustnessstudy:vmagrel4000}. The mean $\delta_{\ell^1}^{(\bm{v})}$ error across ten model runs is \qty{1.43}{\percent}, with a standard deviation $\sigma < \qty{0.09}{\percent}$. Therefore, we can conclude that our proposed model reliably produces highly accurate predictions and is thus well-suited for further investigations.
\begin{figure}[ht]
    \centering
    \input{figures/13_robustness-study/reynolds_4000/rel_error_profile.pgf}
    \caption{Mean value and standard deviation of the normalized error in velocity magnitude obtained from the repeated model training, measured in radial direction along an angle of $\varphi=0$. The errors obtained from the different runs only differ by a very small amount in the outer region close to the wall.}
    \label{fig:nonparameterized:robustnessstudy:vmagrel4000}
\end{figure}


\subsection{Model Configuration}
In our work, we used the \gls{pinn} implementation provided by the \texttt{deepxde} python package \cite{Lu2021DeepXDE} with Tensorflow \cite{Abadi2016TensorFlow} as backend. All models were trained on \num{50000} domain points using minibatch sampling. The optimal size of the \gls{nn} was determined through hyperparameter optimization using the \texttt{optuna} library \cite{Akiba2019Optuna}. The \glspl{nn} employed in \Crefrange{subsec:nonparameterized:baseline}{subsec:nonparameterized:hybridbc} were \num{2} layers deep and \num{100} neurons wide. The parallel \glspl{nn} employed in \Cref{subsec:nonparameterized:domaindecomposed2,subsec:nonparameterized:domaindecomposed3} were \num{25} and \num{100} neurons wide, respectively, and 2 layers deep. We utilized the hyperbolic tangent activation function and applied $\ell^1$ and $\ell^2$ regularization. The loss function was logarithmically transformed, and the L-BFGS-B optimizer \cite{Byrd1995LBFGSTheory,Zhu1997LBFGSBImplementation} from the \texttt{scipy} python package \cite{Virtanen2020SciPy} was employed to perform the minimization. The remaining hyperparameter settings for each model are detailed in \Cref{app:nonparameterized}.

\subsection{Summary}
\label{subsec:nonparameterized:summary}
After presenting the different models, this section aims to equip the reader with the necessary information on which model to choose for their application. The errors and training times for each individual model are reported in \Cref{tab:nonparameterized:summary:models}. All models were trained on the GPUs of the high-performance computer Lichtenberg at the NHR Centers NHR4CES at TU Darmstadt. The reported computing times have been measured there. Note, that the training times generally depend on the available hardware and can differ across different machines. Consequently, they are to be understood more as a general guideline of how expensive the training of a certain model is in comparison to another.

\begin{table}[H]
    \centering
    \caption{Model errors and training times for the \gls{pinn} models discussed in \Cref{sec:nonparameterized}.}
    \label{tab:nonparameterized:summary:models}
    \begin{tabular}{cccccc}
        \toprule
        \multirow{2}{*}{Model} & \multicolumn{2}{c}{$\ell^1$ errors [\unit{\percent}]} & \multicolumn{2}{c}{$\ell^2$ errors [\unit{\percent}]} & \multirow{2}{*}{Time [min]} \\
        \cmidrule(lr){2-3} \cmidrule(lr){4-5} 
        & $\delta^{(\bm{v})}_{\ell^1}$ & $\delta^{(p)}_{\ell^1}$ & $\delta^{(\bm{v})}_{\ell^2}$ & $\delta^{(p)}_{\ell^2}$ & \\
        \midrule
        \makecell{Baseline\\(\Cref{subsec:nonparameterized:baseline})} & \num{10.74} & \num{9.48} & \num{16.41} & \num{14.62} & \num{8.70} \\
        \midrule
        \makecell{Baseline Data\\(\Cref{subsec:nonparameterized:baselinedata})} & \num{2.27} & \num{1.04} & \num{2.93} & \num{1.24} & \num{10.40} \\
        \midrule
        \makecell{Baseline Scaled\\(\Cref{subsec:nonparameterized:baselinescaling})} & \num{3.88} & \num{2.49} & \num{6.47} & \num{4.63} & \num{8.62} \\
        \midrule
        \makecell{Baseline Polar\\(\Cref{subsec:nonparameterized:baselinepolar})} & \num{3.66} & \num{1.14} & \num{4.89} & \num{1.64} & \textbf{\num[text-series-to-math]{5.80}} \\
        \midrule
        \makecell{Strong \gls{bc}\\(\Cref{subsec:nonparameterized:hardbc})} & \num{5.11} & \num{2.26} & \num{7.44} & \num{2.68} & \num{61.93} \\
        \midrule
        \makecell{Hybrid \gls{bc}\\(\Cref{subsec:nonparameterized:hybridbc})} & \num{3.22} & \num{1.20} & \num{4.31} & \num{1.43} & \num{65.08} \\
        \midrule
        \makecell{\gls{dd}\\(\Cref{subsec:nonparameterized:domaindecomposed2})} & \num{1.45} &  \textbf{\num[text-series-to-math]{0.63}} & \num{1.93} & \textbf{\num[text-series-to-math]{0.97}} & \num{51.05} \\
        \midrule
        \makecell{\gls{dd} with outer split\\(\Cref{subsec:nonparameterized:domaindecomposed3})} & \textbf{\num[text-series-to-math]{0.97}} & \num{0.74} & \textbf{\num[text-series-to-math]{1.41}} & \num{1.28} & \num{71.10} \\
        \bottomrule
    \end{tabular}
\end{table}

\Cref{tab:nonparameterized:summary:models} indicates that the model employing \gls{dd} with an additional split of the outer domain presented in \Cref{subsec:nonparameterized:domaindecomposed3} achieves the most accurate predictions for velocity, with $\delta_{\ell^1}^{(\bm{v})} < \qty{1}{\percent}$. As evident from the error distribution in \Cref{fig:nonparameterized:dderr:dd3:vmagrel4000} and the velocity profiles in \Cref{fig:nonparameterized:ddprofile}, this model also performs the best qualitatively in capturing the subtle behavior of the flow. However, the model presented in \Cref{subsec:nonparameterized:domaindecomposed2} performs slightly better in terms of capturing the pressure. Since it is overall less complex and more straightforward than the model defined in \Cref{subsec:nonparameterized:domaindecomposed3}, we chose this model for further work, in particular parameterization by the stirring rate in \Cref{sec:parameterized}. It is apparent from the table that strong \gls{bc} imposition comes at a significant computational cost, increasing the training time by a factor of ${\sim}8$. However, enforcing Dirichlet \glspl{bc} in a strong manner also reduces the number of loss components and thereby the number of hyperparameters that need to be selected. Thus, the time saved in selecting the optimal loss scaling factors should also be considered when comparing the models' training times. Moreover, employing polar coordinates and leveraging the symmetry of the model by only training on $\Omega_\text{sym}$, as described in \Cref{subsec:nonparameterized:baselinepolar}, results in the shortest training time.


\section{Parameterized case}
\label{sec:parameterized}
After identifying the model from \Cref{subsec:nonparameterized:domaindecomposed2} as the best performing model, we parameterize it based on the rotational velocity of the impeller $\omega$ to enable predictions for $\mathrm{Re} \in [\num{1000}, \num{10000}]$. We achieve this by extending the polar domain $\Omega_\text{polar}$ by the parameter space $\Lambda = [\omega_\text{min}, \omega_\text{max}]$, where $\omega_\text{min}$ and $\omega_\text{max}$ for the respective $\mathrm{Re}$ are calculated using \Cref{eq:model:domain:reynolds}. In terms of the notation presented in \Cref{sec:methods}, $\mathcal{X}=\Omega_\text{polar}\times\Lambda$. During the \gls{nn} training, we do not only randomly generate points in $\Omega_\text{polar}$, but additionally draw points from a uniform distribution over $\Lambda$. This section introduces the configuration and results of the parameterized model and subsequently explores the implementation of strong constraints for the domain interface as a means to reduce the number of the model's hyperparameters.


\subsection{\gls{dd} model}
\label{subsec:parameterized:dd}
In this model, the configuration outlined in \Cref{subsec:nonparameterized:domaindecomposed2} was utilized, including output scaling and normalization of $\bm{x}_\text{polar}$. The normalized parametric input, $\tilde{\omega} \in [-1,1]$, was obtained as follows:
\begin{align}
    \label{eq:parameterized:dd:inputscaling}
    \tilde{\omega} &= \frac{\omega - \omega_\text{avg}}{\omega_\text{max}-\omega_\text{avg}} \; , 
\end{align}
where
\begin{align}
    \omega_\text{avg}&=\frac{\omega_\text{min}+\omega_\text{max}}{2} \; .
\end{align}
The position of the domain interface was chosen as $R_\text{inter}=\qty{0.075}{\meter}$. The training of this model took \qty{148.7}{\min}. The distributions of normalized velocity and pressure errors for selected values of $\mathrm{Re}$ are depicted in \Cref{fig:parameterized:dd:violinerr}. It is evident that the model is capable of providing very accurate predictions for velocity and reasonably accurate predictions for pressure within the trained parametric range. Furthermore, parameterizing the model results in an improved mean velocity error for $\mathrm{Re}=\num{4000}$ compared to the non-parameterized model in \Cref{subsec:nonparameterized:domaindecomposed2}, achieving $\delta_{\ell^1}^{(\bm{v})}=\qty{0.9}{\percent}$. 
\begin{figure}[htb]
    \centering
    \begin{subfigure}[t]{0.47\textwidth}
        \centering
        \input{figures/10_DD-polar-2-param/violin_comparison_vel.pgf}
        \caption{Distribution of normalized errors in velocity magnitude for selected values of $\mathrm{Re}$.}
        \label{fig:parameterized:dd:violinerr:velocity}
    \end{subfigure}
    \hfill
    \begin{subfigure}[t]{0.47\textwidth}
        \centering
        \input{figures/10_DD-polar-2-param/violin_comparison_p.pgf}
        \caption{Distribution of normalized errors in pressure for selected values of $\mathrm{Re}$.}
        \label{fig:parameterized:dd:violinerr:pressure}
    \end{subfigure}
    \caption{Distributions of velocity and pressure errors for selected values of $\mathrm{Re}$ of the model parameterized by $\omega$ using soft continuity constraints on the subdomain interface. The ends of the bars indicate minimum and maximum errors, while the additional bar in between represents the mean value, which corresponds to $\delta^{(q)}_{\ell^1}$ in \Cref{eq:model:validation:errormetrics}. The overall performance of the model improves with increasing $\mathrm{Re}$.}
    \label{fig:parameterized:dd:violinerr}
\end{figure}
The relatively larger errors observed for $\mathrm{Re}=\num{1000}$, primarily concentrated in the inner domain near the stirrer tip as illustrated in \Cref{fig:parameterized:dd:reynolds1000err:velerrdistribution}, can be explained by examining the profile of the boundary function $\bm{v}_{\mathrm{BC}}$, used to enforce strong boundary constraints on $\Gamma_\text{stirrer}$ in \Cref{fig:parameterized:dd:reynolds1000err:vbcprofile} (the definition of $\bm{v}_\text{BC}$ can be found in \Cref{app:vbc}). 
While this function offers a good approximation of the kink in velocity at the stirrer tip for $\mathrm{Re}=\num{4000}$, the sharper kink at a lower $\mathrm{Re}$ results in the boundary function being a less accurate representation of the velocity field. Consequently, the \gls{nn} has to make more corrections to this approximation, leading to larger errors. As depicted in \Cref{fig:parameterized:dd:reynolds1000err:velerrdistribution}, another region where higher errors are concentrated is in the vicinity of the baffles and the wall. This can be attributed to the fact that for low $\mathrm{Re}$, the velocity values in this region are very small, making the network more prone to learning the trivial solution, i.e., $\bm{v}=\bm{0}$. The combination of this fact and the previously mentioned characteristics of the velocity profile provides an explanation for why the model's overall performance improves at higher $\mathrm{Re}$. 
\begin{figure}[htb]
    \label{fig:parameterized:dd:reynolds1000err}
    \centering
    \begin{subfigure}[t]{0.47\textwidth}
        \centering
        \input{figures/10_DD-polar-2-param/reynolds_1000/err_vmag_rel_contour.pgf}
        \caption{Spatial distribution of the error in velocity magnitude. Comparatively large errors are localized in the inner domain right at the stirrer tip and in the outer domain near the baffles and reactor walls.}
        \label{fig:parameterized:dd:reynolds1000err:velerrdistribution}
    \end{subfigure}
    \hfill
    \begin{subfigure}[t]{0.47\textwidth}
        \centering
        \input{figures/90_velocity-profiles/vel_profile_bc_phi0.pgf}
        \caption{Velocity profiles along the radius at $\varphi=0$ normalized through division by $\omega$. The reference profile for $\mathrm{Re}=\num{1000}$ exhibits a sharper kink than the one approximated by the boundary function $\bm{v}_\text{BC}$.}
        \label{fig:parameterized:dd:reynolds1000err:vbcprofile}
    \end{subfigure}
    \caption{Sources of error in the velocity predictions made by the parameterized model for $\mathrm{Re}=\num{1000}$. Spatial distribution of velocity error within the domain (A) and comparison of the velocity profile approximated by the boundary function $\bm{v}_\text{BC}$ with profiles of the reference solution for $\mathrm{Re}=\num{1000}$ and $\mathrm{Re}=\num{4000}$ (B).}
\end{figure}

By contrast, pressure predictions are only reasonably accurate. While it might be conceivable to enhance the model's ability to predict pressure through a different selection of loss component scaling factors, exploring all possible combinations becomes impractical due to the extensive parameter space. In the following section, we attempt to address this challenge.


\subsection{\gls{dd} model with domain overlap}
\label{subsec:parameterized:ddoverlap}
As previously demonstrated in \Cref{sec:nonparameterized}, the decision regarding the scaling of individual loss components can profoundly affect the model's prediction capabilities. One of the primary challenges of the model employing \gls{dd} is that, due to the different sets of equations in each subdomain and the additional coupling constraints on $\Gamma_\text{inter}$, its objective function comprises 14 contributions that require scaling. In an effort to reduce the number of loss component scaling factors that need to be chosen, we can employ \Cref{eq:methods:pinn:approximation:strong} to enforce strong constraints for the interface. Following an approach similar to the one outlined in \cite{Moseley2023FBPINNs}, this is achieved by establishing a slight overlap between the two subdomains, as illustrated in \Cref{fig:parameterized:ddoverlap}. Within the overlap region, we augment the output of our \gls{pinn} model in the post-processing step by two auxiliary variables, $v_{\varphi, \text{overlap}}$ and $p_\text{overlap}$, which are defined as follows:
\begin{subequations}
    \label{eq:parameterized:ddoverlap:outputdefinition}
    \begin{align}
        v_{\text{overlap},\varphi}(\bm{x}) &= g_\text{overlap}(\bm{x}) v_{\text{inner},\varphi}(\bm{x}) + (1-g_\text{overlap}(\bm{x})) v_{\text{outer},\varphi}(\bm{x}) \; , \\
        p_\text{overlap}(\bm{x}) &=
        g_\text{overlap}(\bm{x}) p_\text{inner}(\bm{x}) +
        (1-g_\text{overlap}(\bm{x}))
        p_\text{outer}(\bm{x}) \; .
    \end{align}
\end{subequations}
Here $g_\text{overlap}$ is a distance function defined as
\begin{align}
    \begin{cases}
        g_\text{overlap}(\bm{x}) = 0 & \text{on} \; \Gamma_\text{out} \\
         g_\text{overlap}(\bm{x}) = 1 & \text{on} \; \Gamma_\text{in}
    \end{cases} \; ,
\end{align}
which we again determine following the approach proposed in \cite{Sheng2021LFF}.
\begin{figure}[ht]
        \centering

\setlength{\gridlength}{20cm}

\begin{tikzpicture}[font=\large, x=\gridlength, y=\gridlength]
    
    \clip (-0.105,-0.105) rectangle (0.105,0.105);
    
    \def \xThetaP {0.073612159}
    \def \yThetaP {0.0425}
    \def \sqrttwo {1.41421356237}
    \def \position {0.1/\sqrttwo}
    \def \radImp {0.04}
    \def \radIntf {0.065}
    \def \overlap {0.007}
    \def \radInner {\radIntf-\overlap}
    \def \radOuter {\radIntf+\overlap}
    \def \intf {\radIntf/\sqrttwo}
    \def \baffOuter {\position}
    \def \baffInner {0.06}
    
    \fill[very thick, draw=black, fill=custom-blue-1] (0,0) circle (0.1);
    
    \fill[fill=custom-yellow!40] (0,0) circle (\radIntf);
    
    \fill[fill=custom-red!20] (0,0) circle (\radOuter);
    
    \fill[fill=custom-yellow!40] (0,0) circle (\radInner);
    
    \draw[very thick, black] (0,0) circle (0.1);
    
    \draw[very thick, custom-orange, dashed] (0,0) circle (\radIntf);
    
    \draw[very thick, custom-blue-2] (0,0) circle (\radInner);
    
    \draw[very thick, custom-blue-3] (0,0) circle (\radOuter);
    
    \begin{scope}
        \clip (0,0) circle (0.1);
        \def \sqrttwo {1.41421356237}
        \def \position {0.1/\sqrttwo}
        \foreach \x/\y/\d in {\position/\position/-45, \position/-\position/45, -\position/-\position/-45, -\position/\position/45} {
            \node[very thick, draw=black, fill=black!10, shape=rectangle, minimum width=0.005\gridlength, minimum height=0.03\gridlength, inner sep=0pt, anchor=center, rotate=\d] at (\x,\y) {};
        }
    \end{scope}
    
    \foreach \r in {-0.04, 0.04} {
        \draw[very thick] (0,0) -- (0,\r);
        \draw[very thick] (0,0) -- (\r,0);
    }
    
    \node[custom-orange] at (0.032,0.014) {$\Gamma_\text{inter}$};
    \node[custom-blue-2] at (0.025,-0.025) {$\Gamma_\text{in}$};
    \node[custom-blue-3] at (0.0,0.085) {$\Gamma_\text{out}$};
    
\end{tikzpicture}
        \caption{Schematic illustrating the domain overlap defined in \Cref{subsec:parameterized:ddoverlap} and the bounds $\Gamma_\text{in}$ and $\Gamma_\text{out}$ within which the distance function $g_\text{overlap}$ is defined.}
        \label{fig:parameterized:ddoverlap}
\end{figure}
By ensuring the continuity of $v_\varphi$ and $p$ in this manner, we can replace all of the constraints defined in \Cref{eq:nonparameterized:dd:interfacebc} except for \Cref{eq:nonparameterized:dd:interfacebc:vr}, which we still impose weakly, since $v_{\text{inner},r}\equiv0$. Additionally, as shown in \Cref{fig:parameterized:ddoverlap:violinerr}, the accuracy of the model's velocity predictions is comparable to the model defined in \Cref{subsec:parameterized:dd}. Moreover, the present model captures the pressure more accurately. However, it is crucial to note that the increase in training time for this model ranges between $\qty{25}{\percent}$ and $\qty{30}{\percent}$ when compared to the model with weak interface constraints. Nevertheless, this extra time is offset by the spared efforts in selecting the proper scaling factors for the loss components corresponding to continuity constraints. 
\begin{figure}[ht]
    \centering
    \begin{subfigure}[t]{0.47\textwidth}
        \centering
        \input{figures/12_DD-polar-2-param-overlap-pressure/violin_comparison_vel.pgf}
        \caption{Error distribution of velocity magnitude predictions for selected values of $\mathrm{Re}$.}
        \label{fig:parameterized:ddoverlap:violinerr:velocity}
    \end{subfigure}
    \hfill
    \begin{subfigure}[t]{0.47\textwidth}
        \centering
        \input{figures/12_DD-polar-2-param-overlap-pressure/violin_comparison_p.pgf}
        \caption{Error distribution of pressure predictions for selected values of $\mathrm{Re}$.}
        \label{fig:parameterized:ddoverlap:violinerr:pressure}
    \end{subfigure}
    \caption{Velocity and pressure error distributions for selected values of $\mathrm{Re}$ of the model parameterized by $\omega$ with overlapping domains.}
    \label{fig:parameterized:ddoverlap:violinerr}
\end{figure}


\subsection{Summary}
\label{subsec:parameterized:summary}
The model detailed in \Cref{subsec:parameterized:dd} employing weak coupling constraints on $\Gamma_\text{inter}$ achieves a mean velocity error ranging from \qty{0.6}{\percent} to \qty{2}{\percent} across the range $\mathrm{Re} \in [\num{1000}, \num{10000}]$. However, its performance in pressure prediction, with  $\delta_{\ell^1}^{(p)}$ as high as \qty{6}{\percent} is less satisfactory. By enforcing continuity across the domain interface in a strong manner in \Cref{subsec:parameterized:ddoverlap}, we improve the pressure prediction capabilities, achieving $\delta_{\ell^1}^{(p)}$ between \qtyrange{0.3}{1.8}{\percent}, while maintaining the same prediction capabilities for velocity. This improvement is attributed to the elimination of the need to optimize the scaling of loss components representing the velocity and pressure coupling.


\section{Conclusions}
\label{sec:conclusion}

In this paper, we explore various strategies to enhance the performance of \gls{pinn} models for simulating the flow in a 2D \gls{str}. In its most basic configuration, the \gls{pinn} model produces unsatisfactory results for this application case. While integrating a data loss into the training significantly improves prediction capabilities, acquiring such extensive data is often impractical for our application, particularly when considering parameterized models. Therefore, we address the limitations of the baseline \gls{pinn} model without relying on labeled training data. Instead, we incorporate additional insights into the physics of the problem derived from the reference solution. 
We introduce and compare a series of models with varying levels of complexity and accuracy, which we believe can similarly be adapted to different applications. To prove the reproducibility of our results, we examine the robustness of our models and provide detailed reports of the settings of our configurations. 

We show that imposing Dirichlet \glspl{bc} in a strong manner helps to capture kinks in the solution field. Additionally, we observe that the choice of the \gls{bc} function $\bm{h}$ is not arbitrary; the better it represents the physics, the better the prediction capabilities of the model are. Employing \gls{dd} allows us to harness the distinct flow character in different regions of the domain, resulting in both qualitative and quantitative improvements in the model's predictions. Moreover, we showcase that the \gls{xpinn} and \gls{dpinn} approaches can be adapted for cases where different sets of equations are solved in each subdomain, which --- to the best of our knowledge --- has not been demonstrated in the literature before. 

Leveraging the symmetry of the problem and employing polar coordinates yields the shortest training time. Although this model sacrifices some accuracy, it can prove valuable in applications where speed is the top priority. \gls{dd} can be used to train models of the highest accuracy which can serve as basis for detailed analysis of the \gls{str}. 
After showcasing that this model reliably makes highly accurate predictions, we have chosen it for subsequent parameterization by the stirring rate, due to its performance and relative simplicity. 

Through parameterization, we obtain a model achieving a mean velocity error below \qty{2}{\percent} across the range $\mathrm{Re}\in [\num{1000}, \num{10000}]$, but falling short in pressure prediction. We address this limitation by imposing the continuity across the domain interface in a strong manner. By doing so, we reduce the complexity of hyperparameter selection, as we no longer need to optimize the scaling of loss components that account for the coupling of velocity and pressure between the different domains.

Although the training times of the nonparameterized models cannot compete with classical numerical methods, our objective is not to compete in this aspect. Instead, these models serve as preliminary steps for subsequent parameterization. As we demonstrate the interpolation capability of parameterized models within the parameter space, the time required to train one model should be compared with the time needed to conduct numerous numerical simulations. Given that selecting hyperparameters, particularly scaling the loss function components, remains challenging and time-consuming, we mitigate this by enforcing interface continuity constraints strongly, thus reducing the number of these components. 

Our ultimate goal is to apply the methodologies outlined in this work to higher-dimensional models, encompassing both physical and parameter spaces. Moreover, these approaches can be adapted and applied to other complex flow problems beyond the scope of this paper.






\section*{Use of AI tools declaration}
The authors declare they have used Artificial Intelligence (AI) tools in the creation of this article: In some paragraphs throughout the article, AI tools have been used to rephrase and improve sentences grammatically and language-wise. We emphasize that no content has been generated by AI tools.

\section*{Acknowledgments}
This work was performed as part of the Helmholtz School for Data Science in Life, Earth and Energy (HDS-LEE) and received funding from the Helmholtz Association of German Research Centres. This work was supported by the Deutsche Forschungsgemeinschaft (DFG, German Research Foundation) – 333849990/GRK2379 (IRTG Hierarchical and Hybrid Approaches in Modern Inverse Problems).
The authors gratefully acknowledge the computing time provided to them on the high-performance computer Lichtenberg at the NHR Centers NHR4CES at TU Darmstadt. This is funded by the Federal Ministry of Education and Research, and the state governments participating on the basis of the resolutions of the GWK for national high performance computing at universities (\url{www.nhr-verein.de/unsere-partner}). The authors gratefully acknowledge the computing time granted by the JARA Vergabegremium and provided on the JARA Partition part of the supercomputer CLAIX at RWTH Aachen University.

Furthermore the authors want to acknowledge the work of Henri Lubjuhn who developed larger parts of the authors' software framework for conducting the \gls{pinn} trainings during his seminar and master's theses.


\clearpage
\appendix

\section{Supplementary information and hyperparameters for the models presented in
\Cref{sec:nonparameterized}}
\label{app:nonparameterized}


\subsection{Baseline model}
\label{app:nonparameterized:baseline}

The hyperparameters used for the training of the model presented in \Cref{subsec:nonparameterized:baseline} are contained in \Cref{tab:app:nonparameterized:baseline}. The boundary points were fixed throughout all epochs, whereas the domain points were resampled.

\begin{table}[H]
    \centering
    \caption{Hyperparameters of the baseline model presented in \Cref{subsec:nonparameterized:baseline}.}
    \label{tab:app:nonparameterized:baseline}
    \begin{tabular}{lll}
        \toprule
        & Hyperparameter & Value \\
        \midrule
        \multirow{3}{*}{Architecture} & Number of layers & \num{2} \\
        & Number of neurons per layer & \num{100} \\
        & Activation function & $\tanh$ \\
        \midrule
        \multirow{6}{*}{Loss function} & Regularization & $\ell^1+\ell^2$ \\
        & \multirow{5}{*}{Loss scaling} & $\alpha_{\text{momentum},x} = 1$ \\
        & & $\alpha_{\text{momentum},y} = 1$ \\
        & & $\alpha_{\text{mass}} = 1$ \\
        & & $\alpha_\text{wall} = 1$ \\
        & & $\alpha_\text{impeller} = 1$ \\
        \midrule
        \multirow{2}{*}{Optimization} & Optimizer & L-BFGS-B \\
        & Epochs & \num{25000} \\
        \midrule
        \multirow{4}{*}{Sampling} & Number of domain points & \num{2048} \\
        & $\hookrightarrow$ Resampled every & \num{1000} epochs \\
        & \multirow{2}{*}{Number of boundary points} & $\num{1024} \; \text{on} \; \Gamma_\text{stirrer}$ \\
        & & $\num{1024} \; \text{on} \; \Gamma_\text{wall}$ \\
        \bottomrule
    \end{tabular}
\end{table}


\subsection{Baseline model with data}
\label{app:nonparameterized:baselinedata}

The hyperparameters used for the training of the model presented in \Cref{subsec:nonparameterized:baselinedata} are mostly identical with the ones reported in \Cref{tab:app:nonparameterized:baseline}. The additional hyperparameters are listed in \Cref{tab:app:nonparameterized:baselinedata}. The data points were not resampled between the different epochs.

\begin{table}[H]
    \centering
    \caption{Additional hyperparameters of the baseline model with data presented in \Cref{subsec:nonparameterized:baselinedata}.}
    \label{tab:app:nonparameterized:baselinedata}
    \begin{tabular}{lll}
        \toprule
        & Hyperparameter & Value \\
        \midrule
        Loss function & Loss scaling & $\alpha_{\text{data}} = 1$ \\
        \midrule
        Sampling & Number of data points & \num{2000} \\
        \bottomrule
    \end{tabular}
\end{table}


\subsection{Baseline model with loss scaling}
\label{app:nonparameterized:baselinescaling}

The hyperparameters used for the training of the model presented in \Cref{subsec:nonparameterized:baselinescaling} are identical with the ones from \Cref{tab:app:nonparameterized:baseline} except for the loss scaling factors. These have been reported in \Cref{tab:app:nonparameterized:baselinescaling}.

\begin{table}[H]
    \centering
    \caption{Modifications of the hyperparameters of the baseline model with loss scaling presented in \Cref{subsec:nonparameterized:baselinescaling}.}
    \label{tab:app:nonparameterized:baselinescaling}
    \begin{tabular}{lll}
        \toprule
        & Hyperparameter & Value \\
        \midrule
        \multirow{5}{*}{Loss function} & \multirow{5}{*}{Loss scaling} & $\alpha_{\text{momentum},x} = \num{5e1}$ \\
        & & $\alpha_{\text{momentum},y} = \num{5e1}$ \\
        & & $\alpha_{\text{mass}} = \num{5e1}$ \\
        & & $\alpha_{\text{wall}} = \num{5e0}$ \\
        & & $\alpha_{\text{impeller}} = \num{1e0}$ \\
        \bottomrule
    \end{tabular}
\end{table}


\subsection{Baseline model in polar coordinates}
\label{app:nonparameterized:baselinepolar}

The hyperparameters used for the training of the model presented in \Cref{subsec:nonparameterized:baselinepolar} are reported in \Cref{tab:app:nonparameterized:baselinepolar}. Values that are not reported there are identical with the values reported in \Cref{tab:app:nonparameterized:baseline}.

\begin{table}[H]
    \centering
    \caption{Modifications and additions of the hyperparameters of the baseline model in polar coordinates presented in \Cref{subsec:nonparameterized:baselinepolar}.}
    \label{tab:app:nonparameterized:baselinepolar}
    \begin{tabular}{lll}
        \toprule
        & Hyperparameter & Value \\
        \midrule
        \multirow{10}{*}{Loss function} & \multirow{10}{*}{Loss scaling} & $\alpha_{\text{momentum},r} = \num{1e6}$ \\
        & & $\alpha_{\text{momentum},\varphi} = \num{1e6}$ \\
        & & $\alpha_{\text{mass}} = \num{1e0}$ \\
        & & $\alpha_{\text{wall},v_r} = \num{1e1}$ \\
        & & $\alpha_{\text{wall},v_\varphi} = \num{1e1}$ \\
        & & $\alpha_{\text{impeller},v_r} = \num{5e0}$ \\
        & & $\alpha_{\text{impeller},v_\varphi} = \num{5e0}$ \\
        & & $\alpha_{\text{symmetry},v_r} = \num{1e2}$ \\
        & & $\alpha_{\text{symmetry},v_\varphi} = \num{1e2}$ \\
        & & $\alpha_{\text{symmetry},p} = \num{1e0}$ \\
        \midrule
        Optimization & Epochs & \num{12500} \\
        \midrule
        \multirow{5}{*}{Sampling} & Number of domain points & \num{4096} \\ 
        & $\hookrightarrow$ Resampled every & \num{1000} epochs \\
        & \multirow{3}{*}{Number of boundary points} & $\num{512} \; \text{on} \; \Gamma_\text{stirrer}$ \\
        & & $\num{512} \; \text{on} \; \Gamma_\text{wall}$ \\
        & & $\num{1024} \; \text{on} \; \Gamma_\text{sym}$ \\
        \bottomrule
    \end{tabular}
\end{table}


\subsection{Strong \gls{bc} model}
\label{app:nonparameterized:hardbc}

\Cref{tab:app:nonparameterized:hardbc} lists the hyperparameters used for the model presented in \Cref{subsec:nonparameterized:hardbc}. Due to the strong \gls{bc} imposition, no scaling factors for the boundary losses are reported.

\begin{table}[H]
    \centering
    \caption{Hyperparameters of the strong \gls{bc} model presented in \Cref{subsec:nonparameterized:hardbc}.}
    \label{tab:app:nonparameterized:hardbc}
    \begin{tabular}{lll}
        \toprule
        & Hyperparameter & Value \\
        \midrule
        \multirow{3}{*}{Architecture} & Number of layers & \num{2} \\
        & Number of neurons per layer & \num{100} \\
        & Activation function & $\tanh$ \\
        \midrule
        \multirow{4}{*}{Loss function} & Regularization & $\ell^1+\ell^2$ \\
        & \multirow{3}{*}{Loss scaling} & $\alpha_{\text{momentum},x} = \num{1e0}$ \\
        & & $\alpha_{\text{momentum},y} = \num{1e0}$ \\
        & & $\alpha_{\text{mass}} = \num{1e0}$ \\
        \midrule
        \multirow{2}{*}{Optimization} & Optimizer & L-BFGS-B \\
        & Epochs & \num{25000} \\
        \midrule
        \multirow{2}{*}{Sampling} & Number of domain points & \num{2048} \\
        & $\hookrightarrow$ Resampled every & \num{1000} epochs \\
        \bottomrule
    \end{tabular}
\end{table}


\subsection{Hybrid \gls{bc} model}
\label{app:nonparameterized:hybridbc}

For the model with hybrid \glspl{bc} from \Cref{subsec:nonparameterized:hybridbc}, all hyperparameters are identical with the ones reported in \Cref{tab:app:nonparameterized:hardbc}, just one additional scaling factor for the wall \gls{bc} is needed. Additionally, points on $\Gamma_\text{wall}$ need to be sampled. These changes are reported in \Cref{tab:app:nonparameterized:hybridbc}. 

\begin{table}[H]
    \centering
    \caption{Hyperparameters of the hybrid \gls{bc} model presented in \Cref{subsec:nonparameterized:hybridbc}.}
    \label{tab:app:nonparameterized:hybridbc}
    \begin{tabular}{lll}
        \toprule
        & Hyperparameter & Value \\
        \midrule
        Loss function & Loss scaling & $\alpha_{\text{wall}} = \num{1e0}$ \\
        \midrule
        \multirow{1}{*}{Sampling} & Number of boundary points & $\num{1024} \; \text{on} \; \Gamma_\text{wall}$ \\
        \bottomrule
    \end{tabular}
\end{table}


\subsection{\gls{dd} model}
\label{app:nonparameterized:domaindecomposed2}

For the \gls{dd} models, we require significantly more hyperparameters: On the one hand, we now have multiple networks with different architectures and on the other hand, the loss function consists of significantly more terms, consequently resulting in more scaling factors. The hyperparameters for the model presented in \Cref{subsec:nonparameterized:domaindecomposed2} are reported in \Cref{tab:app:nonparameterized:domaindecomposed2}. 

\begin{table}[H]
    \centering
    \caption{Hyperparameters of the \gls{dd} model with two domains presented in \Cref{subsec:nonparameterized:domaindecomposed2}.}
    \label{tab:app:nonparameterized:domaindecomposed2}
    \begin{tabular}{lll}
        \toprule
        & Hyperparameter & Value \\
        \midrule
        \multirow{7}{*}{Architecture} & Network for inner region \\
        & $\hookrightarrow$ Number of layers & \num{2} \\
        & $\hookrightarrow$ Number of neurons per layer & \num{25} \\
        & Network for outer region \\
        & $\hookrightarrow$ Number of layers & \num{2} \\
        & $\hookrightarrow$ Number of neurons per layer & \num{100} \\
        & Activation function & $\tanh$ \\
        \midrule
        \multirow{15}{*}{Loss function} & Regularization & $\ell^1+\ell^2$ \\
        & \multirow{14}{*}{Loss scaling} & $\alpha_{\text{momentum},r} = \num{1e9}$ (inner region) \\
        & & $\alpha_{\text{momentum},\varphi} = \num{1e8}$ (inner region) \\
        & & $\alpha_{\text{momentum},r} = \num{5e13}$ (outer region) \\
        & & $\alpha_{\text{momentum},\varphi} = \num{5e14}$ (outer region) \\
        & & $\alpha_{\text{mass}} = \num{5e8}$ (outer region) \\
        & & $\alpha_{\text{coupling},v_r} = \num{1e12}$ \\
        & & $\alpha_{\text{coupling},v_\varphi} = \num{1e12}$ \\
        & & $\alpha_{\text{coupling},\partial_r v_\varphi} = \num{1e7}$ \\
        & & $\alpha_{\text{coupling},p} = \num{1e6}$ \\
        & & $\alpha_{\text{wall},v_r} = \num{1e10}$ \\
        & & $\alpha_{\text{wall},v_\varphi} = \num{1e13}$ \\
        & & $\alpha_{\text{symmetry},v_r} = \num{1e9}$ \\
        & & $\alpha_{\text{symmetry},v_\varphi} = \num{1e9}$ \\
        & & $\alpha_{\text{symmetry},p} = \num{1e5}$ \\
        \midrule
        \multirow{2}{*}{Optimization} & Optimizer & L-BFGS-B \\
        & Epochs & \num{12500} \\
        \midrule
        \multirow{7}{*}{Sampling} & Number of domain points & \num{4096} \\
        & $\hookrightarrow$ Resampled every & \num{1000} epochs \\
        & $\hookrightarrow$ Position of $R_\text{inter}$ & \num{0.07} \\
        & $\hookrightarrow$ Ratio of points in $\nicefrac{\Omega_\text{inner}}{\Omega_\text{outer}}$ & \num{0.2} \\
         & \multirow{3}{*}{Number of boundary points} & $\num{256} \; \text{on} \; \Gamma_\text{wall}$ \\
        & & $\num{512} \; \text{on} \; \Gamma_\text{sym}$ \\
        & & $\num{256} \; \text{on} \; \Gamma_\text{inter}$ \\
        \bottomrule
    \end{tabular}
\end{table}

Since we are solving the set of \glspl{ode} defined in \Cref{eq:nonparameterized:dd:ode} in $\Omega_\text{inner}$, which is independent of $\varphi$, and we exploit symmetry of $\Omega_\text{outer}$ in the same manner as discussed in \Cref{subsec:nonparameterized:baselinepolar}, the input pre-processing function from \Cref{eq:methods:inputoutputscaling:ansatz} is defined as follows:
\begin{align}
\label{eq:app:nonparameterized:dd:preprocessing}
    \tilde{\bm{x}} &= \begin{pmatrix} r_\text{inner}\\ r_\text{outer}\\ \varphi_\text{outer} \end{pmatrix}= f^\text{pre}(\bm{x}_\text{polar}) = \begin{pmatrix} \nicefrac{(r-R_\text{stirrer})}{(R_\text{inter}-R_\text{stirrer})} \\ \nicefrac{(r+0.5R_\text{reactor}-1.5R_\text{inter})}{0.5(R_\text{reactor}-R_\text{inter})} \\ \nicefrac{4\varphi}{\pi} \end{pmatrix} \; .
\end{align}
The non-dimensional outputs of the \gls{nn} in $\Omega_\text{inner}$ are scaled as:
\begin{align}
    \label{eq:app:nonparameterized:dd:postprocessinginner}
    \bm{u}_\text{inner} = \bm{f}_\text{post}(\bm{\tilde{u}}_\text{inner}) = \begin{pmatrix} \|v_\text{tip}\|\tilde{v}_{\text{inner},\varphi} \\
    \rho v_\text{tip}^2 \tilde{p}_\text{inner}
    \end{pmatrix} \; .
\end{align}
Consequently, the Dirichlet \gls{bc} at $\Gamma_\text{stirrer}$ is imposed in a strong manner by applying \Cref{eq:methods:pinn:approximation:strong} to $v_{\text{inner},\varphi}$. The outputs in $\Omega_\text{outer}$ are scaled as follows:
\begin{align}
    \label{eq:app:nonparameterized:dd:postprocessingouter}
    \bm{u}_\text{outer} = \bm{f}_\text{post}(\bm{\tilde{u}}_\text{outer}) = \begin{pmatrix} v_{\text{norm},r} \tilde{v}_{\text{outer},r} \\
    |v_{\text{norm},\varphi}| \tilde{v}_{\text{outer},\varphi} \\
    \rho\|\bm{v}_\text{norm}\|_2^2 \tilde{p}_\text{outer}
    \end{pmatrix} \; , 
\end{align}
where $v_{\text{norm},r}=v^\text{ref}_{r,\text{max}}$,
\begin{align}
\label{eq:app:nonparameterized:dd:postprocessingouter:vphi}
    v_{\text{norm},\varphi} = v_\text{tip}\frac{R_\text{stirrer}(R_\text{inter}^2-R_\star^2)}{R_\text{inter}(R_\text{stirrer}^2-R_\star^2)} \; ,
\end{align}
and $\bm{v}_\text{norm}=(v_{\text{norm},r},v_{\text{norm},\varphi})^T$. Here, $R_\star=\qty{0.0875}{\meter}$ is chosen as the approximate point where the reference velocity profile depicted in \Cref{fig:app:vbc:velprofiles} hits zero. As the equations solved in $\Omega_\text{inner}$ do not depend on $\varphi$, we only sample points with $\varphi=0$ in this domain.


\newpage

\subsection{\gls{dd} model with outer split}
\label{app:nonparameterized:domaindecomposed3}

For the model with the additional split in the outer domain as introduced in \Cref{subsec:nonparameterized:domaindecomposed3}, the loss function is even further augmented by additional terms. The respective hyperparameters are reported in \Cref{tab:app:nonparameterized:domaindecomposed3}. All values that are not reported coincide with what has been already listed in \Cref{tab:app:nonparameterized:domaindecomposed2}. 

\begin{table}[H]
    \centering
    \caption{Hyperparameters of the \gls{dd} model with outer domain split presented in \Cref{subsec:nonparameterized:domaindecomposed3}.}
    \label{tab:app:nonparameterized:domaindecomposed3}
    \begin{tabular}{lll}
        \toprule
        & Hyperparameter & Value \\
        \midrule
        \multirow{17}{*}{Loss function} & \multirow{17}{*}{Loss scaling} & $\alpha_{\text{momentum},r} = \num{1e11}$ (inner region) \\
        & & $\alpha_{\text{momentum},\varphi} = \num{1e8}$ (inner region) \\
        & & $\alpha_{\text{momentum},r} = \num{4e16}$ (outer region) \\
        & & $\alpha_{\text{momentum},\varphi} = \num{4e16}$ (outer region) \\
        & & $\alpha_{\text{mass}} = \num{4e10}$ (outer region) \\
        & & $\alpha_{\text{coupling},v_r} = \num{1e0}$ \\
        & & $\alpha_{\text{coupling},v_\varphi} = \num{1e14}$ \\
        & & $\alpha_{\text{coupling},\partial_r v_\varphi} = \num{1e8}$ \\
        & & $\alpha_{\text{coupling},p} = \num{1e6}$ \\
        & & $\alpha_{\text{wall},v_r} = \num{1e14}$ \\
        & & $\alpha_{\text{wall},v_\varphi} = \num{1e15}$ \\
        & & $\alpha_{\text{symmetry},v_r} = \num{1e0}$ \\
        & & $\alpha_{\text{symmetry},v_\varphi} = \num{1e10}$ \\
        & & $\alpha_{\text{symmetry},p} = \num{1e5}$ \\
        & & $\alpha_{\text{baffle}} = \num{1e14}$ \\ 
        & & $\alpha_{\text{c}} = \num{1e13}$ \\ 
        & & $\alpha_{\text{d}} = \num{1e8}$ \\ 
        \midrule
        \multirow{9}{*}{Sampling} & Number of domain points & \num{4096} \\
        & $\hookrightarrow$ Resampled every & \num{1000} epochs \\
        & $\hookrightarrow$ Position of $R_\text{inter}$ & \num{0.08} \\
        & $\hookrightarrow$ Ratio of points in $\nicefrac{\Omega_\text{inner}}{\Omega_\text{outer}}$ & \num{0.2} \\
        & \multirow{5}{*}{Number of boundary points} & $\num{528} \; \text{on} \; \Gamma_\text{wall}$ \\
        & & $\num{512} \; \text{on} \; \Gamma_\text{sym}$ \\
        & & $\num{256} \; \text{on} \; \Gamma_\text{inter}$ \\
        & & $\num{256} \; \text{on} \; \Gamma_\text{c}$ \\
        & & $\num{512} \; \text{on} \; \Gamma_\text{baffle}$ \\
        \bottomrule
    \end{tabular}
\end{table}

For this model, the input and output scaling described in \Cref{app:nonparameterized:domaindecomposed2} are applied. The additional output $v_{\text{outer},\varphi}^\text{(2)}$ is scaled by a factor of \num{1e-3}.


\section{Supplementary information and hyperparameters for the models presented \Cref{sec:parameterized}}
\label{app:parameterized}

For both parameterized models outlined in \Cref{subsec:parameterized:dd,subsec:parameterized:ddoverlap} utilizing \gls{dd}, we employ the same \gls{nn} architecture as described in the nonparameterized model in  \Cref{subsec:nonparameterized:domaindecomposed2}. The differences from the nonparameterized model are evident in the selection of loss scaling factors. Additionally, it is crucial to highlight that for both parameterized models, the sampled domain points in $\Omega_\text{inner}$ are not restricted to $\varphi=0$. Instead, sampling is performed for all values of $\varphi \in [0, 2\pi)$ and $\omega \in [\omega_\text{min}, \omega_\text{max}]$.


\subsection{\gls{dd} model}
\label{app:parameterized:dd}

The hyperparameters for the parameterized \gls{dd} model from \Cref{subsec:parameterized:dd} are reported in \Cref{tab:app:parameterized:dd}.

\begin{table}[H]
    \centering
    \caption{Hyperparameters of the parameterized \gls{dd} model with two domains presented in \Cref{subsec:parameterized:dd}.}
    \label{tab:app:parameterized:dd}
    \begin{tabular}{lll}
        \toprule
        & Hyperparameter & Value \\
        \midrule
        \multirow{7}{*}{Architecture} & Network for inner region \\
        & $\hookrightarrow$ Number of layers & \num{2} \\
        & $\hookrightarrow$ Number of neurons per layer & \num{25} \\
        & Network for outer region \\
        & $\hookrightarrow$ Number of layers & \num{2} \\
        & $\hookrightarrow$ Number of neurons per layer & \num{100} \\
        & Activation function & $\tanh$ \\
        \midrule
        \multirow{15}{*}{Loss function} & Regularization & $\ell^1+\ell^2$ \\
        & \multirow{14}{*}{Loss scaling} & $\alpha_{\text{momentum},r} = \num{1e3}$ (inner region) \\
        & & $\alpha_{\text{momentum},\varphi} = \num{1e0}$ (inner region) \\
        & & $\alpha_{\text{momentum},r} = \num{1e7}$ (outer region) \\
        & & $\alpha_{\text{momentum},\varphi} = \num{1e8}$ (outer region) \\
        & & $\alpha_{\text{mass}} = \num{1e3}$ (outer region) \\
        & & $\alpha_{\text{coupling},v_r} = \num{1e7}$ \\
        & & $\alpha_{\text{coupling},v_\varphi} = \num{1e9}$ \\
        & & $\alpha_{\text{coupling},\partial_r v_\varphi} = \num{1e8}$ \\
        & & $\alpha_{\text{coupling},p} = \num{1e3}$ \\
        & & $\alpha_{\text{wall},v_r} = \num{1e2}$ \\
        & & $\alpha_{\text{wall},v_\varphi} = \num{1e4}$ \\
        & & $\alpha_{\text{symmetry},v_r} = \num{1e0}$ \\
        & & $\alpha_{\text{symmetry},v_\varphi} = \num{1e0}$ \\
        & & $\alpha_{\text{symmetry},p} = \num{1e2}$ \\
        \midrule
        \multirow{2}{*}{Optimization} & Optimizer & L-BFGS-B \\
        & Epochs & \num{30000} \\
        \midrule
        \multirow{7}{*}{Sampling} & Number of domain points & \num{4096} \\
        & $\hookrightarrow$ Resampled every & \num{1000} epochs \\
        & $\hookrightarrow$ Position of $R_\text{inter}$ & \num{0.075} \\
        & $\hookrightarrow$ Ratio of points in $\nicefrac{\Omega_\text{inner}}{\Omega_\text{outer}}$ & \num{2.3} \\
        & \multirow{3}{*}{Number of boundary points} & $\num{256} \; \text{on} \; \Gamma_\text{wall}$ \\
        & & $\num{512} \; \text{on} \; \Gamma_\text{sym}$ \\
        & & $\num{256} \; \text{on} \; \Gamma_\text{inter}$ \\
        \bottomrule
    \end{tabular}
\end{table}

The pre- and postprocessing functions defined in \Cref{eq:app:nonparameterized:dd:preprocessing,eq:app:nonparameterized:dd:postprocessinginner,eq:app:nonparameterized:dd:postprocessingouter} are employed. As the scaling for $v_{\text{outer},r}$ in the form of $v_{\text{norm},r}$ now varies with $\omega$, we adapt the scaling accordingly as:
\begin{align}
    \label{eq:app:parameterized:dd:postprocessing:vr}
    v_{\text{norm},r}(\omega) = \num{4e-4}\omega^2 + \num{1.2e-3}\omega \; .
\end{align}


\subsection{\gls{dd} model with domain overlap}
\label{app:parameterized:ddoverlap}
The model with overlapping domains defined in \Cref{subsec:parameterized:ddoverlap} employs the same set of hyperparameters as the model with soft coupling constraints in \Cref{subsec:parameterized:dd}, with the exception of the loss scaling factors. The input and output scaling procedures are also identical. Furthermore, strong coupling constraints are enforced by applying \Cref{eq:parameterized:ddoverlap:outputdefinition} to the scaled outputs and additional points are sampled in the domain overlap. The modifications of the hyperparameters with respect to the ones already reported in \Cref{tab:app:parameterized:dd} are listed in \Cref{tab:app:parameterized:ddoverlap}.

\begin{table}[H]
    \centering
    \caption{Hyperparameters of the parameterized \gls{dd} model with domain overlap presented in \Cref{subsec:parameterized:ddoverlap}.}
    \label{tab:app:parameterized:ddoverlap}
    \begin{tabular}{lll}
        \toprule
        & Hyperparameter & Value \\
        \midrule
        \multirow{10}{*}{Loss function} & \multirow{10}{*}{Loss scaling} & $\alpha_{\text{momentum},r} = \num{1e5}$ (inner region) \\
        & & $\alpha_{\text{momentum},\varphi} = \num{1e0}$ (inner region) \\
        & & $\alpha_{\text{momentum},r} = \num{1e9}$ (outer region) \\
        & & $\alpha_{\text{momentum},\varphi} = \num{1e9}$ (outer region) \\
        & & $\alpha_{\text{mass}} = \num{1e4}$ (outer region) \\
        & & $\alpha_{\text{coupling},v_r} = \num{1e7}$ \\
        & & $\alpha_{\text{wall},v_r} = \num{1e4}$ \\
        & & $\alpha_{\text{wall},v_\varphi} = \num{1e4}$ \\
        & & $\alpha_{\text{symmetry},v_r} = \num{1e0}$ \\
        & & $\alpha_{\text{symmetry},v_\varphi} = \num{1e0}$ \\
        & & $\alpha_{\text{symmetry},p} = \num{1e2}$ \\
        \midrule
        \multirow{5}{*}{Sampling} & Number of domain points & \num{4096} \\
        & $\hookrightarrow$ Resampled every & \num{1000} epochs \\
        & $\hookrightarrow$ Ratio of points in $\nicefrac{\Omega_\text{inner}}{\Omega_\text{outer}}$ & \num{0.125} \\ 
        & Number of domain overlap points & \num{1536} \\
        & $\hookrightarrow$ Width of overlap & \num{0.01} \\
        \bottomrule
    \end{tabular}
\end{table}


\section{Boundary function approximating the flow field}
\label{app:vbc}

In the following we derive the functions $\tilde{\bm{v}}_{\text{BC}}$ and $\bm{v}_{\text{BC}}$ that are used to strongly impose the \glspl{bc} of our model problem for the models presented in \Cref{subsec:nonparameterized:hardbc,subsec:nonparameterized:hybridbc,subsec:nonparameterized:domaindecomposed2,subsec:nonparameterized:domaindecomposed3}. For more convenient notation, the functions are derived in polar coordinates. 

From \Cref{eq:model:domain:bcs} it follows that
\begin{align}
    v_r = 0 \quad \text{on} \quad \Gamma_\text{stirrer} \cup \Gamma_\text{wall} \; .
\end{align}
We can therefore set $\tilde{v}_{\text{BC},r}=v_{\text{BC},r}=0$ in the whole domain to satisfy all Dirichlet \glspl{bc}. 

As pointed out in \Cref{subsec:nonparameterized:hardbc}, our findings indicate that the choice of the boundary functions greatly influences the model predictions. Consequently, we aim to define $\tilde{v}_{\text{BC},\varphi}$ such that it approximates the reference velocity profile illustrated in \Cref{fig:app:vbc:velprofiles}.
\begin{figure}[ht]
    \centering
    \input{figures/90_velocity-profiles/vel_profile_bc_lff_phi0.pgf}
    \caption{Reference velocity profile along the radius at $\varphi=0$ for $\mathrm{Re}=\num{4000}$ and the function $\bm{v}_\text{BC}$ constructed to approximate it.}
    \label{fig:app:vbc:velprofiles}
\end{figure}
We can observe that the behavior of $v^\text{ref}_\varphi$ distinctly differs between the impeller region $r\leq R_\text{stirrer}$ and the outer region $R_\text{stirrer}<r<R_\text{reactor}$. This motivates a piece-wise definition of $\tilde{v}_{\text{BC},\varphi}$, which increases linearly in the stirrer region and then decays with unknown slope in the outer region. Since the increase in the inner region is given by the \gls{bc} \Cref{eq:model:domain:bcs:stirrer}, we have
\begin{align}
    \label{eq:vbc:unscaled}
    \tilde{v}_{\text{BC},\varphi} &= \begin{cases}
        \omega r & r\leq R_\text{stirrer} \\
        \tilde{v}_{\text{BC},\text{outer},\varphi}(r) & R_\text{stirrer}<r 
    \end{cases} \; ,
\end{align}
where $\tilde{v}_{\text{BC},\text{outer},\varphi}$ models the decay of the reference solution observed in \Cref{fig:app:vbc:velprofiles} for $r>R_\text{stirrer}$. In order to obtain a continuous profile, we know that \\$\tilde{v}_{\text{BC},\text{outer},\varphi}(R_\text{stirrer})=\omega r$. Moreover, we observe that the profile flattens out and approaches $0$ at some $R^\star\gg R_\text{stirrer}$, i.e., $\tilde{v}_{\text{BC},\text{outer},\varphi}(R_\star)=0$. As detailed in \Cref{subsec:nonparameterized:domaindecomposed2}, the governing equations can be approximated by an \gls{ode} system under some simplifying assumptions. Equipped with the conditions just mentioned, i.e., 
\begin{subequations}
\begin{align}
    \tilde{v}_{\text{BC},\text{outer},\varphi}(r)&=\omega r && \text{at} \; r=R_\text{stirrer} \; , \\
    \tilde{v}_{\text{BC},\text{outer},\varphi}(r)&=0 && \text{at} \; r=R_\star \; ,
\end{align}
\end{subequations}
\Cref{eq:nonparameterized:dd:ode:vphi} can be solved analytically and we choose its solution as $\tilde{v}_{\text{BC},\text{outer},\varphi}$, i.e., we have
\begin{align}
    \tilde{v}_{\text{BC},\text{outer},\varphi}(r) = \omega R_\text{stirrer}\frac{R_\text{stirrer}(r^2-R_\star^2)}{r(R_\text{stirrer}^2-R_\star^2)} \; ,
\end{align}
where $R_\star$ is a tuning parameter to adjust the slope of the velocity profile. For our application, we chose a value of $R_\star=\qty{0.0875}{\meter}$ which is approximately the location where the reference velocity profile depicted in \Cref{fig:app:vbc:velprofiles} flattens out.

We should note, that $\tilde{v}_{\text{BC},\varphi}$ defined in \Cref{eq:vbc:unscaled} does not fulfill the \gls{bc} on the wall as illustrated in \Cref{fig:app:vbc:velprofiles}. While this is sufficient for the models with hybrid boundary constraints where we only strongly enforce the \gls{bc} on the impeller, it is not suitable for the model presented in \Cref{subsec:nonparameterized:hardbc}. Therefore, we propose an additional weighting inspired by the construction of the boundary function $\bm{g}(\bm{x})$ in \cite{Sheng2021LFF}, which reads
\begin{align}
    s(r) &= \left(1-[1-l(r)]^\mu\right) \; ,
\end{align}
where $l(r)$ is a spline constructed following the approach outlined in \cite{Sheng2021LFF} to fulfill the condition
\begin{align}
    \begin{cases}
        l(r)=0 & \text{on} \; \Gamma_\text{wall} \\
        l(r)>0 & \text{in} \; \Omega_\text{polar} \setminus\Gamma_\text{wall} \\
    \end{cases} \; ,
\end{align}
and the hyperparameter $\mu=8$ is set for this application. With that, a $v_{\text{BC},\varphi}$ which also fulfills the wall \gls{bc} is obtained as 
\begin{align}
    \label{eq:vbc:scaled}
    v_{\text{BC},\varphi}(r) &= s(r) \tilde{v}_{\text{BC},\varphi}(r) \; .
\end{align}

\textit{Remark:} As pointed out in the beginning of this section, we report the boundary functions in polar coordinates as it simplifies their derivation. In order to employ them for the models presented in \Cref{subsec:nonparameterized:hardbc,subsec:nonparameterized:hybridbc} which use Cartesian coordinates as input, the functions have to be composed with a mapping from Cartesian to polar coordinates accordingly.









\bibliographystyle{AIMS}
\bibliography{literature.bib}

\medskip
Received xxxx 20xx; revised xxxx 20xx; early access xxxx 20xx.
\medskip

\end{document}